\documentclass[useAMS,usenatbib]{mn2e}
\usepackage{graphicx,amssymb,times}

\title[7-mm spectral imaging of Sgr~B2]{Spectral imaging of the 
Sagittarius~B2 region in multiple 7-mm molecular lines}
\author[P. A. Jones et al.]{P. A. Jones$^{1,2}$
\thanks{E-mail:pjones@phys.unsw.edu.au},
M. G. Burton$^{1}$, N. F. H. Tothill$^{1,3}$ and M. R. Cunningham$^{1,2}$ 
\\
$^{1}$School of Physics, University of New South Wales, Sydney, NSW 2052, 
Australia \\
$^{2}$Departamento de Astronom\'\i a, Universidad de Chile, Casilla 36-D, 
Santiago, Chile \\
$^{3}$School of Physics, University of Exeter, Stocker Road, Exeter, 
EX 4 4 QL, UK
}

\begin{document}

\date{Accepted 2010 October 11. Received 2010 October 8; in original form 
2010 July 14}

\pagerange{\pageref{firstpage}--\pageref{lastpage}} \pubyear{2010}

\maketitle

\label{firstpage}

\begin{abstract}

We have undertaken a spectral-line imaging 
survey of a $6 \times 6$ arcmin$^2$ area around Sgr~B2 
near the centre of the Galaxy, in the 
range from 30 to 50 GHz, using the Mopra telescope. 
The spatial resolution varies from 
1.0 to 1.4 arcmin
and the spectral resolution from 1.6 to 2.7 km~s$^{-1}$ over the frequency range.
We present velocity-integrated 
emission images for 47 lines: 38 molecular lines and 9 radio 
recombination lines.

There are significant differences between the distributions of different
molecules, in part due to spatial differences in chemical abundance across
the complex. For example, HNCO and HOCO$^{+}$ are found preferentially in
the north cloud, and CH$_{2}$NH near Sgr B2 (N). Some of the differences
between lines are due to excitation differences, as shown by 
the 36.17 and 44.07 GHz lines of CH$_{3}$OH, which have maser emission,
compared to the 48.37 GHz line of CH$_{3}$OH. Other major differences
in integrated molecular line distribution are due to absorption of the
7-mm free-free continuum emission (spatially traced by the radio recombination
line emission) by cool intervening molecular material, causing a central dip
in the molecular line distributions.

These line distribution similarities and differences have been statistically
described by principal component analysis (PCA), and interpreted in terms
of simple Sgr B2 physical components of the cooler, lower density envelope,
and dense, hot cores Sgr B2 (N), (M) and (S).

\end{abstract}


\section{Introduction}

Sagittarius B2 (Sgr~B2) is a giant
molecular cloud complex near the centre of the Galaxy.
There is spectacular star-formation activity, much of it deeply embedded
in molecular cores, as traced by the high far-infrared luminosity,
compact and ultracompact H {\sc ii} regions \citep{ga+95}, with
maser emission from water \citep*{mcgode04}, hydroxyl \citep{gacl90},
formaldehyde \citep{megopa94} and methanol, both class I and class II 
\citep{ca96,meme97}. 

Three major star-forming centres are located in a 
north-south line, labelled north, middle and south, Sgr~B2 (N), (M) and (S),
which have strong continuum radio H {\sc ii} region free-free emission,
and submillimetre and millimetre wavelength thermal dust emission 
\citep{go+93,pi+00}. These cores are surrounded by a larger, less dense
envelope.

The hot cores Sgr~B2 (N) and (M) are particularly rich in molecules,
and so have been the targets of many millimetre spectral-line surveys 
\citep{culith86,tu89,nu+98,nu+00,be+05,be+07}. The PRIMOS
survey with the Greenbank
Telescope (GBT) will survey Sgr~B2 (N) over the whole band from 300~MHz to 
50~GHz \citep{re+08}. 

Sgr B2 is one of the most prominent features of the Central Molecular Zone
(CMZ), the bar-shaped \citep{sa+04} molecular region in the central few 
hundred pc of the Galaxy, as shown in emission of $^{12}$CO, $^{13}$CO
\citep{ok+98}, CS \citep*{tshauk99} and HNCO \citep{da+97}, for example.

The distance to Sgr B2 has been measured as $7.8^{+0.8}_{-0.7}$~kpc by 
\citet{re+09}, with the proper motion suggesting that Sgr B2 is nearer by 
0.13~kpc than the Galactic centre, and the projected distance from the 
Galactic centre is 0.09~kpc. For calculations, we assume a distance to Sgr 
B2 of $8.0$~kpc.

The Sgr B2 molecular cloud has complex kinematics, with the densest core
centred on Sgr B2 (M) at velocity 60 - 65 km~s$^{-1}$, and this area
corresponding to a `hole' in the emission in CO and CS at 40 - 50 km~s$^{-1}$
\citep{sa+00}. This has been interpreted by \citet{ha+94,ha+08} as a cloud-cloud 
collision between a clump at 70 - 80 km~s$^{-1}$, and the 40 - 50 km~s$^{-1}$ 
cloud triggering the current star formation: the collision changes the velocity
of material, leaving a deficit (the hole) in the 40 - 50 km~s$^{-1}$ range,
as material shifts to the intermediate velocity of the 60 - 65 km~s$^{-1}$ range.

The spatial distribution in the Sgr B2 molecular cloud is also complex,
but can be largely described by a ridge of emission in a north-south line
near the Sgr~B2 (N), (M) and (S) hot cores, or centred to the west of these 
cores \citep{jo+08}. This ridge continues to a peak 1 arcmin north of 
Sgr~B2 (N), the `north cloud' which is enhanced in HNCO and HOCO$^{+}$
\citep{mi+98}, possibly associated with the shock of the cloud-cloud collision.
This ridge of emission coincides with the Sgr B2 extended envelope, as traced 
by sub-mm dust emission \citep{pi+00}. 

In this paper, we present results of imaging the Sgr~B2 region 
($6 \times 6$~arcmin$^{2}$) in multiple spectral lines, 
over 30 to 50 GHz (the 7-mm band), using the Mopra telescope.
This follows on a similar spectral-line imaging survey of Sgr~B2 in the 
3-mm band, presented as \citet{jo+08}, \citet*{jobucu08} and \citet*{jobulo08}.
This study provides spatially resolved line distributions of a large number
of molecules and transitions to constrain chemical models of the Sgr B2 complex.

\section[]{Observations and Data Reduction}

The observations were made with the 22-m Mopra radio telescope, of the 
Australia Telescope National Facility (ATNF) using the 
UNSW-MOPS digital filterbank.
The Mopra MMIC receiver has a bandwidth of 8 GHz, and the MOPS backend can
cover the full 8-GHz range in the broad band mode. This gives 
four 2.2-GHz sub-bands each with 8192 channels of 0.27 MHz. There is also a 
zoom mode with up to sixteen spectra (up to four in each 2.2-GHz sub-band) 
of width 137 MHz. This provides higher spectral resolution
with 4096 channels of 0.033 MHz in each zoom spectrum. We chose the 
broad band mode, as the lines in Sgr~B2 are wide, so that the 0.27 MHz 
channels, corresponding to around 1.6 to 2.7 km~s$^{-1}$, are quite adequate
to resolve their structure.
This mode also allowed a complete line-mapping survey to be made, without 
pre-selecting
which lines would be covered, or the restrictions of selecting at most four 
lines in each 2.2 GHz sub-band.

The Mopra receiver covers the range 30 to 50 GHz in the so-called 7-mm band, 
so the band actually covers wavelengths 6 to 10 mm. We chose three
tunings centred at 45.7, 38.0 and 34.1 GHz, to cover the range whole 
30 to 50 GHz range. The latter two tunings overlap, with two 2.2 GHz sub-bands
in common in the range 34 to 38 GHz.

The area was observed with on-the-fly (OTF) mapping, in an area
$6 \times 6$ arcmin$^{2}$, centred near Sgr~B2~(N) and (M), 
in a similar way to that described in \citet{jo+08}. However, unlike the 3-mm
imaging of \citet{jo+08}, we used galactic coordinates for the scan direction, 
and hence the square scan area. We used position switching for bandpass calibration 
with the off-source reference position observed before each 6 arcmin long source
scan. 
The system temperature was calibrated with a noise diode. The 7-mm Mopra system
does not use paddle scans (unlike the 3-mm system), and hence does not correct
for the absorption through the atmosphere. This effect on calibration is around
the 10 percent level, so we take that as the major uncertainty in the data. 

The OTF data were turned into FITS data cubes with the {\sc livedata} and
{\sc gridzilla} 
packages\footnote{http://www.atnf.csiro.au/people/mcalabre/livedata.html}. 
The raw data files in 
RPFITS\footnote{http://www.atnf.csiro.au/computing/software/rpfits.html} 
format,
were bandpass corrected using the off-source reference spectra with 
{\sc livedata},
a robust second order polynomial fitted to the baseline and subtracted, and 
the data output
as SDFITS \citep{ga00} spectra. These spectra were then gridded into 
datacubes using
{\sc gridzilla}, with a median filter for the interpolation. The median was 
used, as this is much more robust to the effect of bad data. 

The whole 8-GHz band for each tuning was first turned into four cubes
with frequency pixels for the 2.2-GHz sub-bands, making twelve cubes overall
covering the entire 30 to 50 GHz range. These were used to conduct a full line 
survey in this spectral range. We compiled a list of the strongest (identified) 
lines, as given in Table \ref{tab:lines_table}.

The {\sc gridzilla}
scripts also allowed the lines (Table \ref{tab:lines_table}) to be specfied,
with their rest frequencies, so that the {\sc gridzilla} output provides
FITS cubes with velocity coordinates. We used this for a second pass through
the data, to produce individual cubes for each line, over a velocity 
range -300 to 300 km~s$^{-1}$. For the data over the frequency range 34 to 38 GHz,
where two of the tunings overlapped, we combined the data from both 
observations to improve the signal-to-noise.

The FITS cubes were then read into the {\sc miriad} \citep{satewr95} 
package for further processing and analysis. As the emission is typically
of low surface brightness, the data were smoothed in velocity, with a 3-point
hanning kernel, to make data cubes with improved surface
brightness sensitivity. This gives around 0.54 MHz, or 3.2 to 5.4 km~s$^{-1}$ 
effective FWHM spectral resolution (across the spectral range), and ensures 
the data are Nyquist sampled in velocity.

The resolution of the Mopra beam in the 7-mm band was recently measured by 
\cite{ur+10} to vary between 0.99~arcmin at 49~GHz and 1.37~arcmin at 31~GHz,
or roughly $\theta = \lambda/D$ (where $D$ is the 22-m dish diameter). 
We will assume these values to give resolution between 1.0 arcmin at 50 GHz,
and 1.4 arcmin at 30 GHz, with the caveats that: a) this is somewhat smaller 
than previously quoted in the Mopra documentation as 82 arcsec (1.37 arcmin) at 
42 GHz\footnote{http://www.narrabri.atnf.csiro.au/mopra/obsinfo.html}, and;
b) actually gives a somewhat flatter variation than expected from 
$\theta \propto \lambda$.
Since we are mostly concerned in this paper 
with the spatial and velocity structure, we have left the intensities 
in the T$_{A}^*$ scale,
without correction for the main beam efficiency onto the T$_{MB}$ scale.
The measured main beam efficiencies $\eta_{MB}$ vary between 43 and 53~percent
within the range 
31 to 49 GHz, and extended beam efficiencies $\eta_{XB}$ from 
56 to 69~percent \citep{ur+10}.

\section[]{Results}
\label{sec:results}

We have produced data cubes for the 47 strongest lines in the 30 to 50 GHz range,
as listed in Table \ref{tab:lines_table}. The flag in the last column of
Table \ref{tab:lines_table} refers to the source of the line identification.
Most are from the NIST 
database\footnote{http://physics.nist.gov/PhysRefData/Micro/Html/contents.html}
of lines known in the interstellar medium \citep{lodr04}. As Sgr~B2 is among 
the richest known 
sources of interstellar lines, these lines are mostly already 
well-known in Sgr~B2.
There are also some radio recombination lines, taken from the splatalogue 
compilation\footnote{http://www.splatalogue.net/}. We have identified several
lines in the JPL 
database\footnote{http://spec.jpl.nasa.gov/ftp/pub/catalog/catform.html}
\citep{pi+98}
corresponding to other transitions of molecules (CH$_{2}$NH, NH$_{2}$CHO, 
NH$_{2}$CN, CH$_{3}$CHO) known in Sgr~B2 from other frequencies,
notably in the 3-mm band in \citet{jo+08}.

The RMS noise in the (Hanning smoothed) data cubes varies from 34 to 76 mK
(T$_{A}^*$).

There are also a few other weaker lines 
detected (mostly at the Sgr~B2 (N) and (M) cores), but as they are quite weak 
and we do not have confident line 
identifications, we not consider them further here. Other projects are in 
progress to obtain line survey data for Sgr~B2, notably the 
PRIMOS\footnote{http://www.cv.nrao.edu/ aremijan/PRIMOS} survey with the 100-m 
Green Bank Telescope (GBT) of Sgr~B2~(N) which will cover this 7-mm band 
\citep{re+08}.
These line surveys have longer integration time and higher sensitivity, and
are better for studying the weak lines in the cores than the mapping 
observations presented here.

\begin{table}
\caption{Summary of strong lines detected in the Mopra
observations. The first column gives an approximate frequency we have used
for labelling transitions in this paper. The second column identifies the
species, and the third its transition.
The last columns indicates the rest frequency and the database used for the 
identification: L, Lovas: S, splatalogue; J, JPL (see text). The rest 
frequencies with a * indicate the frequency used for lines
corresponding to multiple transitions, most also 
indicated as group (gp) in the transition list.}
\begin{tabular}{ccccc}
\hline
Approx.     & Line ID          &                    & Exact      &     \\
Freq.     & molecule         & Transition         & Rest Freq. &     \\
GHz       & or atom          &                    & MHz        &     \\
\hline
30.00  &  SO            &  1(0)~--~0(1)           &  30001.547  & L  \\
31.22  &  H             &  59 $\alpha$            &  31223.313  & S  \\
31.95  &  HC$_{5}$N     &  12~--~11  	          &  31951.777  & L  \\ 
32.85  &  H             &  58 $\alpha$            &  32852.196  & S  \\
33.75  &  CCS           &  3,2~--~2,1             &  33751.370  & L  \\
34.60  &  H             &  57 $\alpha$            &  34596.383  & S  \\
34.61  &  HC$_{5}$N     &  13~--~12               &  34614.385  & L  \\     
35.07  &  CH$_{2}$NH    &  3(0,3)~--~2(1,2) gp    &  35065.545* & J  \\
35.27  &  H$^{13}$CCCN  &  4~--~3 group           &  35267.440* & L  \\  
36.17  &  CH$_{3}$OH    &  4($-$1,4)~--~3(0,3) E  &  36169.290  & L  \\
36.39  &  HC$_{3}$N     &  4~--~3 group           &  36392.365* & L  \\ 
36.47  &  H             &  56 $\alpha$            &  36466.26   & S  \\
36.49  &  OCS           &  3~--~2                 &  36488.813  & L  \\    
36.80  &  CH$_{3}$CN    &  2(0)~--~1(0) gp        &  36795.568* & L  \\
37.28  &  HC$_{5}$N     &  14~--~13               &  37276.985  & L  \\           
38.47  &  H             &  55 $\alpha$            &  38473.358  & S  \\
38.506 &  CH$_{3}$CHO   &  2(0,2)~--~1(0,1) E     &  38506.035  & L  \\
38.512 &  CH$_{3}$CHO   &  2(0,2)~--~1(0,1) A     &  38512.081  & L  \\                
39.36  &  CH$_{3}$CHO   &  2(1,1)~--~1(1,0) E     &  39362.533  & J  \\  
39.59  &  CH$_{3}$CHO   &  2(1,1)~--~1(1,0) A     &  39594.292  & J  \\       
39.73  &  NH$_{2}$CN    &  2(1,2)~--~1(1,1),v=0   &  39725.3811 & J  \\
39.94  &  HC$_{5}$N     &  15~--~14               &  39939.574  & L  \\     
40.25  &  CH$_{2}$CN    &  2~--~1 group           &  40253.884* & L \\
40.63  &  H             &  54 $\alpha$            &  40630.498  & S  \\
40.88  &  NH$_{2}$CHO   &  2(1,2)~--~1(1,1) gp    &  40875.2766* & J  \\ 
42.39  &  NH$_{2}$CHO   &  2(0,2)~--~1(0,1) gp    &  42386.070* & J  \\  
42.60  &  HC$_{5}$N     &  16~--~15  	          &  42602.153  & L  \\
42.67  &  HCS$^+$       &  1~--~0                 &  42674.197  & L  \\
42.77  &  HOCO$^+$      &  2(0,2)~--~1(0,1)       &  42766.1975 & L  \\ 
42.88  &  $^{29}$SiO    &  1~--~0 v=0             &  42879.922  & L  \\
42.95  &  H             &  53 $\alpha$            &  42951.968  & S  \\      
43.42  &  SiO           &  1~--~0 v=0  	          &  43423.864  & L  \\
43.96  &  HNCO          &  2(0,2)~--~1(0,1) gp    &  43962.998* & L  \\
44.07  &  CH$_{3}$OH    &  7(0,7)~--~6(1,6) A+    &  44069.476  & L  \\       
44.08  &  H$^{13}$CCCN  &  5~--~4                 &  44084.172  & L  \\    
45.26  &  HC$_{5}$N     &  17~--~16               &  45264.720  & L  \\     
45.30  &  HC$^{13}$CCN  &  5~--~4                 &  45297.346  & L  \\         
       &  HCC$^{13}$CN  &  5~--~4                 &  45301.707* & L  \\
45.38  &  CCS           &  4,3~--~3,2             &  45379.029  & L  \\
45.45  &  H             &  52 $\alpha$            &  45453.719  & S  \\
45.49  &  HC$_{3}$N     &  5~--~4 group           &  45490.316* & L  \\
46.25  &  $^{13}$CS     &  1~--~0                 &  46247.580  & L  \\
47.93  &  HC$_{5}$N     &  18~--~17               &  47927.275  & L  \\      
48.15  &  H             &  51 $\alpha$            &  48153.597  & S  \\
48.21  &  C$^{34}$S     &  1~--~0                 &  48206.946  & L  \\
48.37  &  CH$_{3}$OH    &  1(0,1)~--~0(0,0) A+    &  48372.467* & L  \\      
       &  CH$_{3}$OH    &  1(0,1)~--~0(0,0) E     &  48376.889  & L  \\
48.65  &  OCS           &  4~--~3                 &  48651.6043 & L  \\              
48.99  &  CS            &  1~--~0                 &  48990.957  & L  \\
\hline
\end{tabular}
\label{tab:lines_table}
\end{table}

\begin{table}
\caption{Summary of peaks fitted to the lines. The lines are grouped into
the same order as plotted in Figs. \ref{fig:RRLs_int} and 
\ref{fig:HC3N+HC5N_int} to \ref{fig:misc_int}.
The positions are the fits to the maximum in the channel of strongest emission,
and the velocity centre, velocity width (FWHM) and peak are from gaussian 
fits at the brightest pixel of the line emission. Note that the velocity
centres use the rest frequencies from Table \ref{tab:lines_table} so
will be confused for the groups of lines, and the velocity widths will be
increased for such confused lines.}
\label{tab:fitted_peaks}
\begin{tabular}{ccccccc}
\hline
 Line ID    &  Rest   & Lat.  & Long. & Veloc. & Veloc. & Peak \\
 molecule   &  Freq.  & peak  & peak  & centre & width  & $T_{A}^*$ \\
 or atom    &  GHz    &  deg  & deg   & \multicolumn{2}{c}{km\,s$^{-1}$} & K \\
\hline
H 59 $\alpha$ & 31.22 & 0.675 & -0.031 & 62 & 48 & 0.40  \\
H 58 $\alpha$ & 32.85 & 0.674 & -0.034 & 65 & 38 & 0.45  \\
H 57 $\alpha$ & 34.60 & 0.672 & -0.035 & 64 & 34 & 0.46  \\
H 56 $\alpha$ & 36.47 & 0.676 & -0.034 & 65 & 36 & 0.40  \\
H 55 $\alpha$ & 38.47 & 0.674 & -0.035 & 65 & 34 & 0.43  \\
H 54 $\alpha$ & 40.63 & 0.672 & -0.036 & 62 & 46 & 0.40  \\
H 53 $\alpha$ & 42.95 & 0.670 & -0.034 & 67 & 35 & 0.33  \\
H 52 $\alpha$ & 45.45 & 0.672 & -0.034 & 63 & 32 & 0.33  \\
H 51 $\alpha$ & 48.15 & 0.675 & -0.034 & 64 & 54 & 0.39  \\
              &       &       &        &    &    &       \\
HC$_{3}$N     & 36.39 & 0.692 & -0.021 & 65 & 31 & 2.58  \\
HC$_{3}$N     & 45.49 & 0.691 & -0.022 & 66 & 30 & 2.17  \\
HCC$^{13}$CN $^{a}$ & 45.30 & 0.682 & -0.018 & 84 & 43 & 0.17  \\
H$^{13}$CCCN  & 35.27 & 0.694 & -0.010 & 65 & 25 & 0.20  \\
H$^{13}$CCCN  & 44.08 & 0.679 & -0.019 & 69 & 26 & 0.19  \\  
HC$_{5}$N     & 31.95 & 0.699 & -0.022 & 65 & 24 & 0.29  \\
HC$_{5}$N     & 34.61 & 0.691 & -0.015 & 65 & 21 & 0.33  \\
HC$_{5}$N     & 37.28 & 0.685 & -0.010 & 67 & 22 & 0.31  \\
HC$_{5}$N     & 39.94 & 0.675 & -0.015 & 68 & 21 & 0.34  \\
HC$_{5}$N     & 42.60 & 0.688 & -0.021 & 69 & 20 & 0.32  \\
HC$_{5}$N     & 45.26 & 0.698 & -0.024 & 64 & 21 & 0.27  \\
HC$_{5}$N     & 47.93 & 0.696 & -0.025 & 65 & 25 & 0.39  \\
              &       &       &        &    &    &       \\
CH$_{3}$OH    & 36.17 & 0.693 & -0.024 & 67 & 22 & 35.7  \\
CH$_{3}$OH    & 44.07 & 0.672 & -0.028 & 65 & 21 & 1.77  \\
CH$_{3}$OH    & 48.37 & 0.689 & -0.022 & 64 & 35 & 2.59  \\
CH$_{3}$CHO   & 38.506 & 0.687 & -0.014 & 70 & 26 & 0.26  \\
CH$_{3}$CHO   & 38.512 & 0.679 & -0.007 & 64 & 39 & 0.22  \\
CH$_{3}$CHO   & 39.36 & 0.687 & -0.021 & 70 & 18 & 0.16  \\
CH$_{3}$CHO   & 39.59 & 0.684 & -0.012 & 68 & 20 & 0.21  \\
NH$_{2}$CHO   & 40.88 & 0.684 & -0.008 & 64 & 18 & 0.21  \\
NH$_{2}$CHO   & 42.39 & 0.681 & -0.012 & 70 & 25 & 0.24  \\
              &       &       &        &    &    &       \\
SO $^{b}$     & 30.00 & 0.695 & -0.008 & 58 & (25) & 0.26 \\
SiO           & 43.42 & 0.707 & -0.021 & 57 & 50 & 0.67  \\
$^{29}$SiO    & 42.88 & 0.673 & -0.009 & 67 & 25 & 0.14  \\
CS $^{c}   $  & 48.99 & 0.664 & -0.040 & 51 & 11 & 1.91  \\
              &       &       &        & 83 & 26 & 1.39  \\
$^{13}$CS     & 46.25 & 0.656 & -0.041 & 60 & 25 & 0.31  \\
C$^{34}$S     & 48.21 & 0.662 & -0.042 & 61 & 33 & 0.45  \\
CCS           & 33.75 & 0.693 & -0.020 & 66 & 24 & 0.27  \\
CCS           & 45.38 & 0.683 & -0.008 & 65 & 18 & 0.24  \\
CH$_{3}$CN    & 36.80 & 0.680 & -0.012 & 72 & 33 & 0.71  \\
              &       &       &        &    &    &       \\
OCS           & 36.49 & 0.676 & -0.032 & 63 & 20 & 0.47  \\
OCS           & 48.65 & 0.692 & -0.024 & 62 & 25 & 0.52  \\
HNCO          & 43.96 & 0.696 & -0.021 & 64 & 24 & 2.30  \\
HOCO$^+$      & 42.77 & 0.682 & -0.011 & 67 & 20 & 0.34  \\
CH$_{2}$NH    & 35.07 & 0.668 & -0.034 & 60 & 20 & 0.48  \\
NH$_{2}$CN    & 39.73 & 0.699 & -0.024 & 67 & 47 & 0.20  \\
HCS$^+$       & 42.67 & 0.674 & -0.011 & 66 & 19 & 0.17  \\
CH$_{2}$CN    & 40.25 & 0.651 & -0.055 & 54 & 69 & 0.21  \\
\hline
\end{tabular}
$^{a}$ line labelled as HCC$^{13}$CN a blend with HC$^{13}$CCN so 
larger line width and biased velocity \\
$^{b}$ SO line at band edge, fitted with fixed width, due to poor 
data \\
$^{c}$ CS line with self-absorption, fitted as two gaussians
\end{table}

To study the spatial distribution of the different lines, we have made 
integrated emission images (Figs. \ref{fig:RRLs_int} and \ref{fig:HC3N+HC5N_int}
to \ref{fig:misc_int}).
We integrated each data cube over a velocity range  which included significant 
emission (above around 3 $\sigma$ in the cubes) for each line, which means the 
velocity range is different for each line.

The integrated images in Figs. \ref{fig:RRLs_int} and \ref{fig:HC3N+HC5N_int} 
to \ref{fig:misc_int}
are plotted as grey-scale, with the scale bar to the right in K~km~s$^{-1}$ 
($T_{A}^{*}$). The contour levels are in equal linear steps, mostly 
2 K~km~s$^{-1}$, but sometimes 5 or 10 K~km~s$^{-1}$, and 100 K~km~s$^{-1}$
for the very strong CH$_{3}$OH maser line at 36.17 GHz. We plot as fiducial
marks the positions of radio sources as crosses and mid-IR sources as squares,
as in \citet{jo+08}. As the axes of Figs. \ref{fig:RRLs_int} to \ref{fig:misc_int}
are Galactic coordinates, the north-south line
of cores Sgr~B2~(N), (M) and (S) is at a (Galactic) position angle 
around 45 degrees, with Sgr~B2~(N) being the cross (radio source) near the 
centre at $l = 0.680$~deg, $b = -0.028$~deg, Sgr~B2~(M) being the cross and 
square (radio and infrared source) below and to the right at 
$l = 0.667$~deg, $b = -0.036$~deg and  Sgr~B2~(S) the cross and 
square further below and to the right at 
$l = 0.657$~deg, $b = -0.041$~deg.
The radio and infrared peaks are plotted
(with the N, M and S labels) in Fig. \ref{fig:radio_submm_ir} in equatorial 
coordinates. 

We discuss these images in more detail below, in subsections \ref{subs:rrls}
to \ref{subs:misc}.

For quantitative analysis of the differences in spatial distribution of the
different lines, we have fitted the position of the peak emission in the data
cubes, with the {\sc miriad} task {\it maxfit}, which fits the position with a 
parabolic fit to the spatial pixels around the peak pixel in the cube. 
This is similar to the analysis used in \citet{jo+08} for the 3-mm Sgr~B2 Mopra
imaging, but we only fit the strongest peak, not multiple peaks, as the 
resolution is lower here at 7 mm and much of the spatial structure is merged 
together. As the emission is typically complex and extended, we do not 
generally fit the spatial structure as gaussian peaks.

These peak positions are tabulated in Table 
\ref{tab:fitted_peaks}, arranged in groups of similar lines, as plotted in 
Figs. \ref{fig:RRLs_int} and \ref{fig:HC3N+HC5N_int} to \ref{fig:misc_int}. 
We note that due to
absorption in some of the spectra, and line width differences across the area, 
that the position of the peak emission in the data cubes is not the same as the 
position of the peak of integrated emission.

The velocity structure was studied by fitting the spectrum in each data cube
at the position of the pixel of strongest emission, with the {\sc miriad} task 
{\it gaufit}. The fitted velocity and velocity width are given in Table 
\ref{tab:fitted_peaks}. The peak brightness ($T_{A}^{*}$) at this position
is also listed, to give an indication of line strength. We have used
the hanning smoothed data cubes, with spectral resolution 3.2 to 5.4 
km~s$^{-1}$, for the fitting, but this is not expected to significantly
affect the velocity widths.

\subsection{Hydrogen recombination lines}
\label{subs:rrls}

There are nine radio recombination lines (RRLs) of hydrogen, in the series 
H~51~$\alpha$ to H~59~$\alpha$, shown in Fig. \ref{fig:RRLs_int}. They peak 
between 
the two strong free-free radio continuum sources Sgr~B2~(N) and (M), with
both sources contributing to the RRL emission. The mean peak position for the
nine lines (Table \ref{tab:fitted_peaks}) is  $l = 0.673$~deg, 
$b = -0.034$~deg, 
mean velocity 64~km~s$^{-1}$ and mean velocity width 40~km~s$^{-1}$.

We show in Fig. \ref{fig:RRL_7mm_overlay} the peak positions of the RRLs
relative to (our previously unpublished) 
higher resolution Australia Telescope Compact Array (ATCA) 
observations of the continuum. The low resolution Mopra data merge together
the RRL emission from Sgr~B2~(N) and (M). We also note (see the Appendix)
that the 7-mm free-free emission has an extended diffuse component,
in addition to the cores Sgr~B2~(N) and (M), which is resolved out
in the ATCA observations. The extended component contributes close to half
of the radio flux in the 30 to 50~GHz range.

For comparison BIMA observations of H~59~$\alpha$ RRL in both 
Sgr~B2~(M) and (N) are given in \citet*{pelisn00} and 
higher resolution VLA observations of the H~52~$\alpha$ in Sgr~B2~(M) 
are given by \citet{de+96}.

The 7-mm free-free emission probed by the RRLs and shown in
Fig. \ref{fig:RRL_7mm_overlay} also indicates where there may be absorption 
of molecular lines (Figs. \ref{fig:HC3N+HC5N_int} to \ref{fig:misc_int}) by the 
strong Sgr~B2~(N) and (M) continuum.

\begin{figure*}
\includegraphics[width = 5.2 cm,angle=-90]{SgrB2_2009_7mm_RRL_H59a_int3.ps}
\includegraphics[width = 5.2 cm,angle=-90]{SgrB2_2009_7mm_RRL_H58a_int3.ps}
\includegraphics[width = 5.2 cm,angle=-90]{SgrB2_2009_7mm_RRL_H57a_int3.ps}
\includegraphics[width = 5.2 cm,angle=-90]{SgrB2_2009_7mm_RRL_H56a_int3.ps}
\includegraphics[width = 5.2 cm,angle=-90]{SgrB2_2009_7mm_RRL_H55a_int3.ps}
\includegraphics[width = 5.2 cm,angle=-90]{SgrB2_2009_7mm_RRL_H54a_int3.ps}
\includegraphics[width = 5.2 cm,angle=-90]{SgrB2_2009_7mm_RRL_H53a_int3.ps}
\includegraphics[width = 5.2 cm,angle=-90]{SgrB2_2009_7mm_RRL_H52a_int3.ps}
\includegraphics[width = 5.2 cm,angle=-90]{SgrB2_2009_7mm_RRL_H51a_int3.ps}
\caption{The integrated emission images for the nine recombination lines of 
hydrogen H 59 $\alpha$ to H 51 $\alpha$. In this figure and later images
the crosses (+) denote the position of radio peaks and the open squares 
($\Box$) positions of infrared peaks. The line
of cores Sgr~B2~(N), (M) and (S) is at a (Galactic) position angle 
around 45 degrees, with Sgr~B2~(N) being the cross near the 
centre at $l = 0.680$~deg, $b = -0.028$~deg, Sgr~B2~(M) being the cross and 
square below and to the right at 
$l = 0.667$~deg, $b = -0.036$~deg and  Sgr~B2~(S) the cross and 
square further below and to the right at 
$l = 0.657$~deg, $b = -0.041$~deg. For clarity, these cores are labelled
in this figure, and later figures. The scale bar is in units of 
K~km~s$_{-1}$, for $\int T_{A}^{*} dv$.}
\label{fig:RRLs_int}
\end{figure*}

\begin{figure}
\includegraphics[height = 8.5 cm,angle=-90]{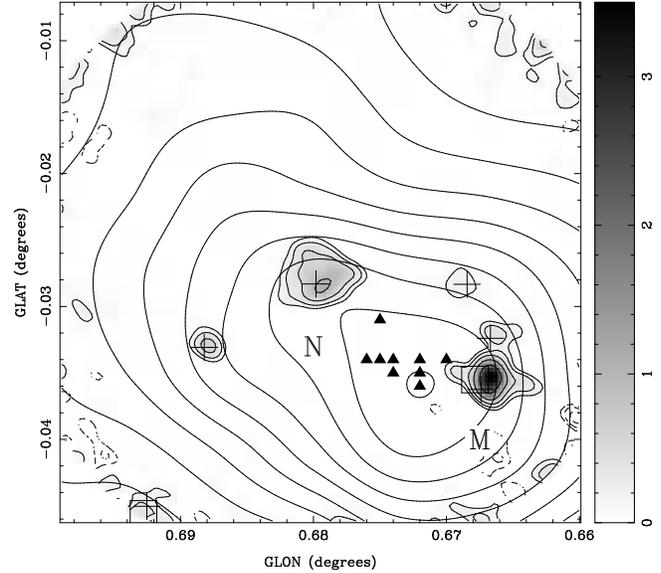} 
\caption{The fitted peak positions of the 7-mm radio recombination lines 
(filled) triangles, relative to the high resolution 7-mm radio continuum 
emission, from ATCA observations, in grey-scale and contours, with the low 
resolution Mopra integrated H~51~$\alpha$ also plotted as contours. The 
recombination lines include emission from Sgr~B2~(N), centre, Sgr~B2~(M),
lower right, and the extended envelope which is resolved out in the ATCA data.}
\label{fig:RRL_7mm_overlay}
\end{figure}

\subsection{HC$_{3}$N and $^{13}$C isotopologues}
\label{subs:hc3n}

\begin{figure*}
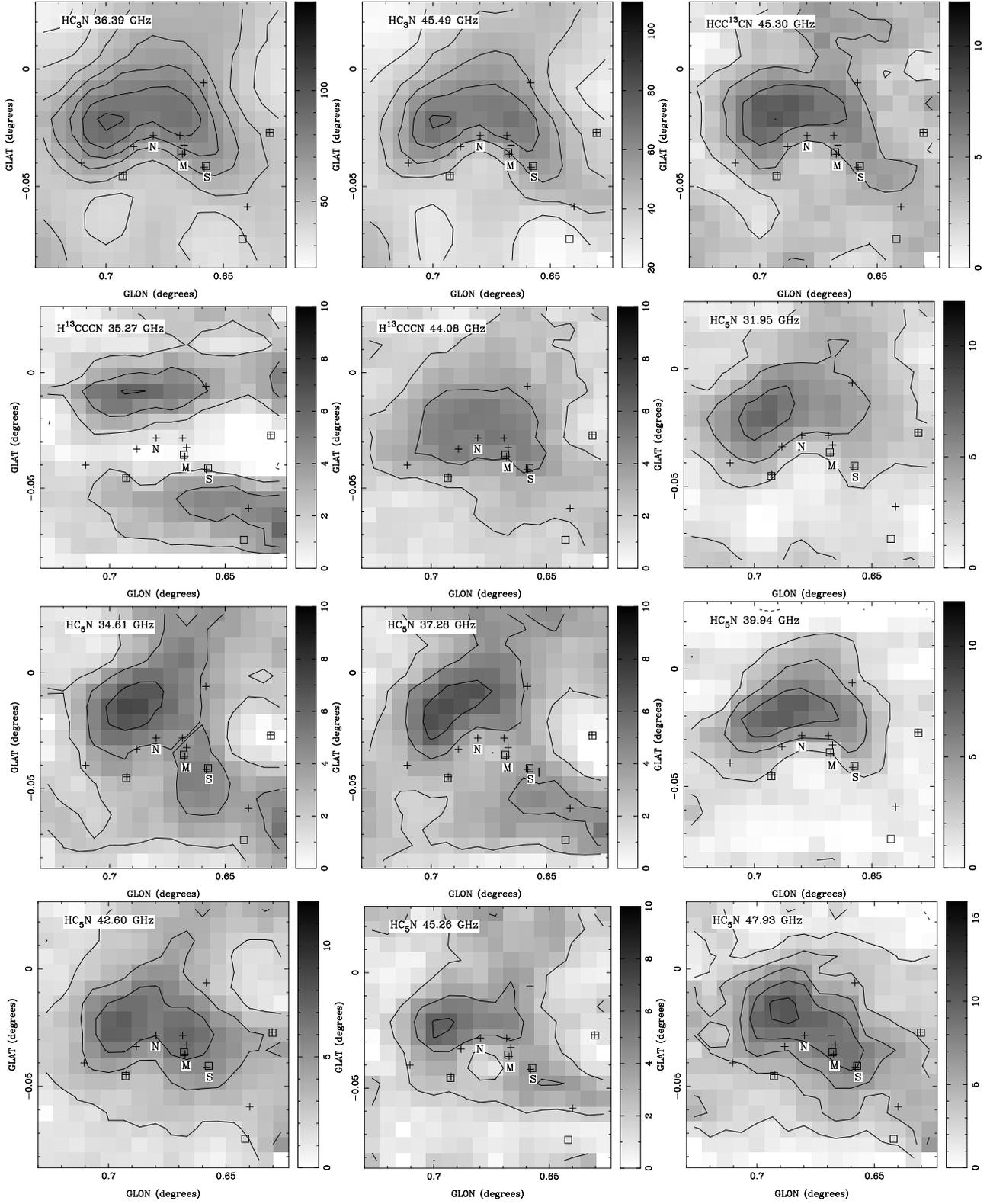

\includegraphics[width = 5.2 cm,angle=-90]{SgrB2_2009_7mm_HCCCN_B_int3.ps}
\includegraphics[width = 5.2 cm,angle=-90]{SgrB2_2009_7mm_HCCCN_A_int3.ps}
\includegraphics[width = 5.2 cm,angle=-90]{SgrB2_2009_7mm_HCC13CN_int3.ps}
\includegraphics[width = 5.2 cm,angle=-90]{SgrB2_2009_7mm_H13CCCN_B_int3.ps}
\includegraphics[width = 5.2 cm,angle=-90]{SgrB2_2009_7mm_H13CCCN_A_int3.ps}
\includegraphics[width = 5.2 cm,angle=-90]{SgrB2_2009_7mm_HC5N_G_int3.ps}
\includegraphics[width = 5.2 cm,angle=-90]{SgrB2_2009_7mm_HC5N_E_int3.ps}
\includegraphics[width = 5.2 cm,angle=-90]{SgrB2_2009_7mm_HC5N_D_int3.ps}
\includegraphics[width = 5.2 cm,angle=-90]{SgrB2_2009_7mm_HC5N_F_int3.ps}
\includegraphics[width = 5.2 cm,angle=-90]{SgrB2_2009_7mm_HC5N_A_int3.ps}
\includegraphics[width = 5.2 cm,angle=-90]{SgrB2_2009_7mm_HC5N_B_int3.ps}
\includegraphics[width = 5.2 cm,angle=-90]{SgrB2_2009_7mm_HC5N_C_int3.ps}
\caption{The integrated emission images for two lines of HC$_{3}$N, 
three lines of $^{13}$C isotopologues of HC$_{3}$N, and seven lines of
HC$_{5}$N.}
\label{fig:HC3N+HC5N_int}
\end{figure*}

There are two lines of cyanoacetylene HC$_{3}$N, shown in Fig. 
\ref{fig:HC3N+HC5N_int}, the 4--3 (36.39 GHz) and 5--4 (45.49 GHz) groups
of transitions. The integrated emission is distributed on a ridge
at higher latitudes than the Sgr~B2~(N), (M) and (S) line (shown as crosses),
that is to the north-west in equatorial coordinates, with the peak to the north
of Sgr~B2~(N). This is similar to the distribution of the HC$_{3}$N 3-mm 
lines 9--8, 10--9, 11--10 and 12--11 found in \citet{jo+08}.
It is also similar to the distribution of the 4--3 line in \citet{ligo91}
and \citet*{chohmo94}, but these latter data covered smaller areas. 

There are also three lines of the $^{13}$C isotopologues of HC$_{3}$N, the
combined HCC$^{13}$CN and HCC$^{13}$CN 5--4 line (45.30 GHz), the 
H$^{13}$CCCN 4--3 (35.27 GHz) and H$^{13}$CCCN 5--4 (44.08 GHz) lines.
These are all quite weak lines, but the 45.30 GHz line shows good agreement
in spatial distribution to the HC$_{3}$N lines. The other two show fair
agreement, but may be affected by scanning artifacts, with stripes along the
longitude direction.

The mean peak position for the five lines is  $l = 0.688$ deg, 
$b = -0.018$~deg, and, for the four lines not blended,
mean velocity 66~km~s$^{-1}$ and mean velocity width 28~km~s$^{-1}$.
(Note that the blended 45.30 GHz line has larger width, and anomalous velocity,
in Table \ref{tab:fitted_peaks}.)

\subsection{HC$_{5}$N}
\label{subs:hc5n}

There are seven lines of cyanodiacetylene HC$_{5}$N, shown in Fig. 
\ref{fig:HC3N+HC5N_int}, the series from 12--11 (31.95 GHz) to 18--17 
(47.93 GHz) transitions. The lines are all quite weak, but the distribution
is similar to that of HC$_{3}$N (above), best seen in the 31.95 GHz and 39.94
GHz lines, but broadly similar in the other lines, with the ridge to more
positive latitudes than the radio sources (crosses).
The mean peak position for the seven lines is  $l = 0.690$~deg, 
$b = -0.019$~deg,
mean velocity 66~km~s$^{-1}$ and mean velocity width 22~km~s$^{-1}$.

\subsection{CH$_{3}$OH}
\label{subs:ch3oh}

\begin{figure*}
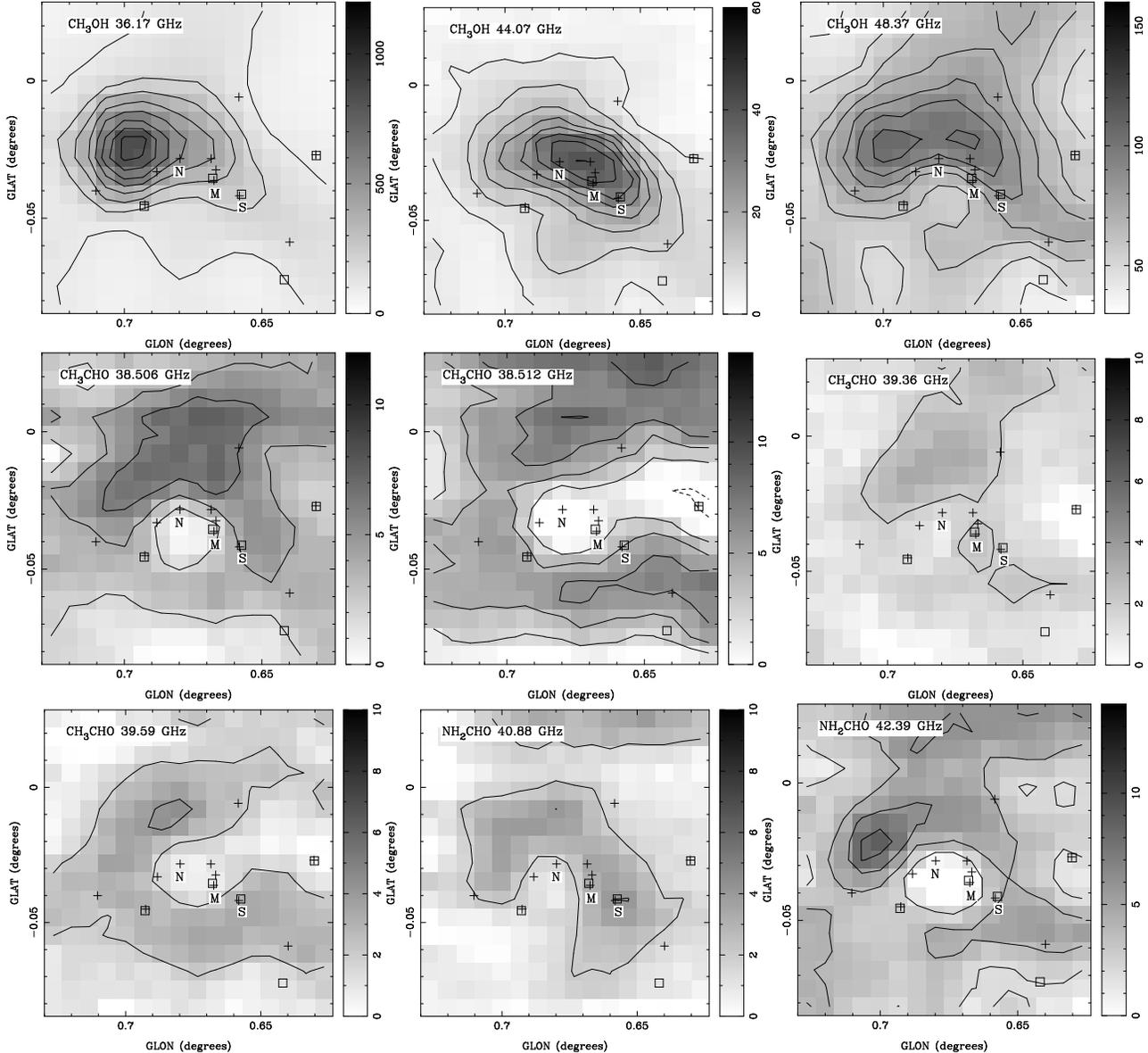

\includegraphics[width = 5.2 cm,angle=-90]{SgrB2_2009_7mm_CH3OH_C_int3.ps}
\includegraphics[width = 5.2 cm,angle=-90]{SgrB2_2009_7mm_CH3OH_A_int3.ps}
\includegraphics[width = 5.2 cm,angle=-90]{SgrB2_2009_7mm_CH3OH_B_int3.ps}
\includegraphics[width = 5.2 cm,angle=-90]{SgrB2_2009_7mm_CH3CHO_D_int3.ps}
\includegraphics[width = 5.2 cm,angle=-90]{SgrB2_2009_7mm_CH3CHO_A_int3.ps}
\includegraphics[width = 5.2 cm,angle=-90]{SgrB2_2009_7mm_CH3CHO_B_int3.ps}
\includegraphics[width = 5.2 cm,angle=-90]{SgrB2_2009_7mm_CH3CHO_C_int3.ps}
\includegraphics[width = 5.2 cm,angle=-90]{SgrB2_2009_7mm_NH2CHO_B_int3.ps}
\includegraphics[width = 5.2 cm,angle=-90]{SgrB2_2009_7mm_NH2CHO_A_int3.ps}
\caption{The integrated emission images for three lines of CH$_{3}$OH 
(methanol), four lines of CH$_{3}$CHO and two lines of NH$_{2}$CHO.}
\label{fig:CH3OH_etc_int}
\end{figure*}

There are three lines of methanol CH$_{3}$OH, shown in Fig. 
\ref{fig:CH3OH_etc_int}, the 4($-$1,4)~--~3(0,3) E (36.17 GHz),
 7(0,7)~--~6(1,6) A+ (44.07 GHz) and the 1(0,1)~--~0(0,0) A+  plus
1(0,1)~--~0(0,0) E (48.37 GHz) lines. The three line have quite different
distributions, due to the complex collisional and radiative excitation
processes of methanol. 

The 36.17 GHz line can act as a (class I) maser, which gives
the high brightness peak seen here in  Fig. 
\ref{fig:CH3OH_etc_int}. Note that the peak is at the north cloud 
\citep{jo+08} at $l = 0.693$~deg, $b = -0.024$~deg.
The distribution is in good agreement with that found
by \citet{liwi96} for this line, with higher resolution Effelsberg 100-m 
telescope observations. 

The 44.07 GHz line can also act as a (class I) maser.
\citet{meme97} have used the VLA at much higher resolution than the Mopra data 
here to map the distribution of this line. They find many maser and
quasi-thermal peaks near
the Sgr~B2~(N) and (M) cores. The Mopra data (Fig. \ref{fig:CH3OH_etc_int})
are consistent with this distribution.

The 48.37 GHz line has a distribution that is similar to HC$_{3}$N 
(Fig. \ref{fig:HC3N+HC5N_int}) and other 3-mm transitions of methanol
in \citet{jo+08}. This is not a maser transition  so comparison
with the 44.07 GHz line is instructive \citep{va+91}.

\subsection{CH$_{3}$CHO}
\label{subs:ch3cho}

There are four lines of acetaldehyde CH$_{3}$CHO, shown in Fig. 
\ref{fig:CH3OH_etc_int}, the 2(0,2)~--~1(0,1) E (38.506 GHz), 2(0,2)~--~1(0,1) A
(38.512 GHz), 2(1,1)~--~1(1,0) E (39.36 GHz) and 2(1,1)~--~1(1,0) A (39.59 GHz)
lines.

The distribution of integrated emission from acetaldehyde is widespread,
as found by \citet{chka03}. However, we find for these 7-mm lines that there 
is a deficit of emission near Sgr~B2~(N) and (M), probably due to absorption.
\citet{chka03} find the 1(1,1)~--~1(1,0) transition at 1065 MHz in 
emission near Sgr~B2~(N), (M) and (S), which they attribute to a weak maser.
The distribution of the four lines is similar, although the 39.36 and 39.59 GHz
lines are weak. The emission peaks at the north cloud with 
mean peak position $l = 0.684$ deg, $b = -0.014$~deg, 
mean velocity 68~km~s$^{-1}$ and mean velocity width 26~km~s$^{-1}$.

\subsection{NH$_{2}$CHO}
\label{subs:nh2cho}

There are two lines of formamide NH$_{2}$CHO, shown in Fig. 
\ref{fig:CH3OH_etc_int}, the 2(1,2)~--~1(1,1) group (40.88 GHz) and the
2(0,2)~--~1(0,1) group (42.39 GHz). The distribution of integrated emission
is similar to that of acetaldehyde (above) with widespread emission,
peaking at the north cloud (mean peak position $l = 0.683$ deg, 
$b = -0.010$~deg) and a deficit near Sgr~B2~(N) and (M).
This distribution is somewhat different to that of the formamide 
5(1,4)~--~4(1,4) line at 102.07 GHz in \citet{jo+08} which peaks
on the ridge to the west of Sgr~B2~(M), but this line is confused
with the protonated formaldehyde H$_{2}$COH$^{+}$ 4(0,4)~--~3(1,3) line
\citep{oh+96}.

\subsection{SO}
\label{subs:so}

\begin{figure*}
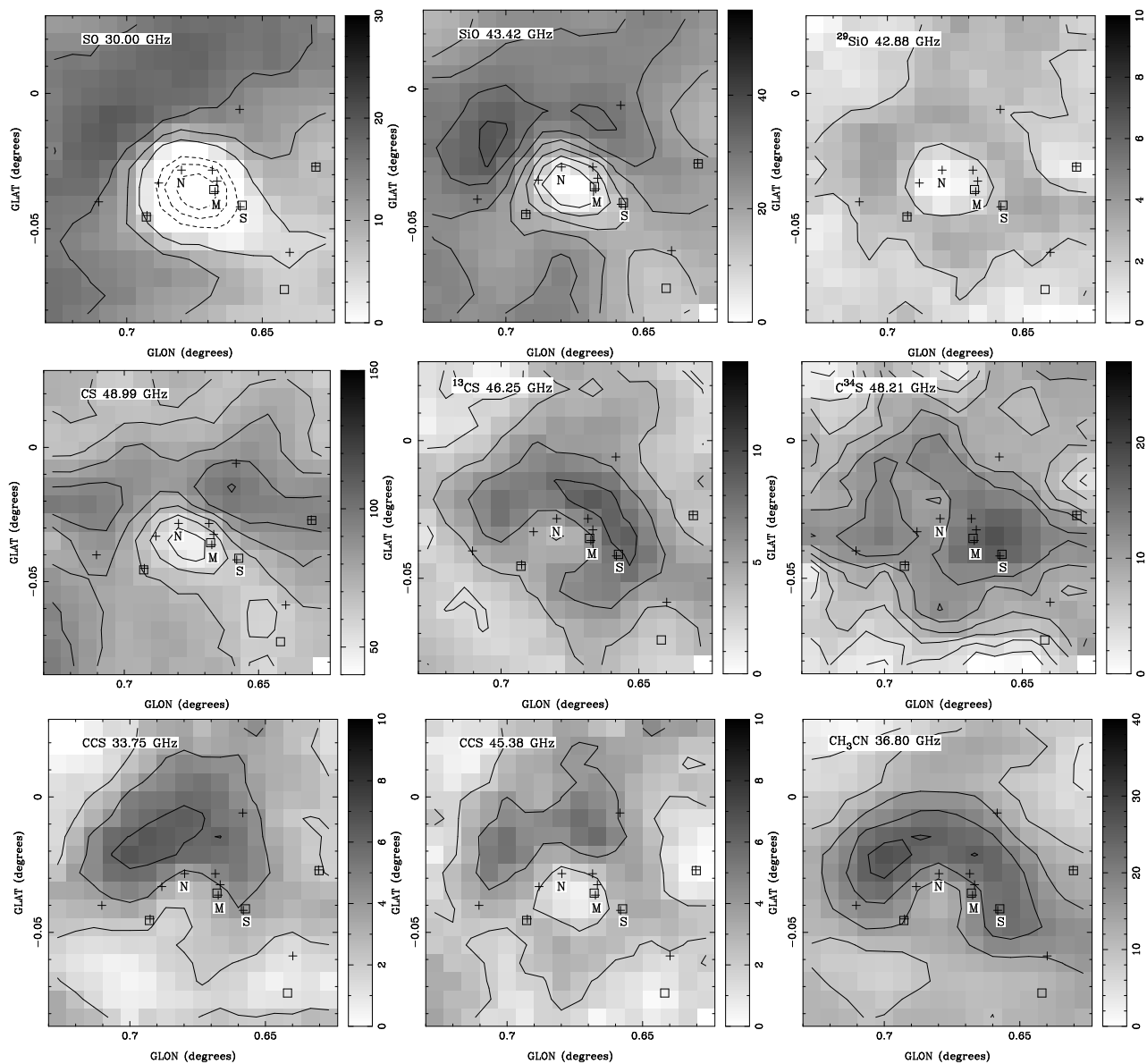

\includegraphics[width = 5.2 cm,angle=-90]{SgrB2_2009_7mm_SO_nar_int3.ps}
\includegraphics[width = 5.2 cm,angle=-90]{SgrB2_2009_7mm_SiO_int3.ps}
\includegraphics[width = 5.2 cm,angle=-90]{SgrB2_2009_7mm_29SiO_int3.ps}
\includegraphics[width = 5.2 cm,angle=-90]{SgrB2_2009_7mm_CS_int3.ps}
\includegraphics[width = 5.2 cm,angle=-90]{SgrB2_2009_7mm_13CS_int3.ps}
\includegraphics[width = 5.2 cm,angle=-90]{SgrB2_2009_7mm_C34S_int3.ps}
\includegraphics[width = 5.2 cm,angle=-90]{SgrB2_2009_7mm_CCS_B_int3.ps}
\includegraphics[width = 5.2 cm,angle=-90]{SgrB2_2009_7mm_CCS_A_int3.ps}
\includegraphics[width = 5.2 cm,angle=-90]{SgrB2_2009_7mm_CH3CN_int3.ps}
\caption{The integrated emission images for one line each of SO,
SiO, $^{29}$SiO, CS, $^{13}$CS and C$^{34}$S,
two lines of CCS and one line of CH$_{3}$CN. Note that the dashed contours 
indicate negative integrated emission, or absorption.}
\label{fig:SO_etc_int}
\end{figure*}

There is one line of sulphur monoxide SO, shown in Fig. 
\ref{fig:SO_etc_int}, the  1(0)~--~0(1) (30.00 GHz) line.
There is widespread emission over the area, with strong absorption
near Sgr~B2~(M). This is in contrast to the 3-mm SO lines, where 
the emission is strongly peaked at Sgr~B2~(M) in the 2(2)~--~1(1) (86.09 GHz)
and  2(3)~--~1(2) (109.25 GHz) lines, and extended along the ridge-line
to the west of  Sgr~B2~(M) in the 3(2)~--~2(1) (99.30 GHz) line
\citep{jo+08,go+87}.  


\subsection{SiO and $^{29}$SiO}
\label{subs:sio}

There are two lines of silicon monoxide SiO 1~--~0 v=0 (43.42 GHz),
and the isotopologue $^{29}$SiO 1~--~0 v=0 (42.88 GHz) 
shown in Fig. \ref{fig:SO_etc_int}. The integrated emission is widely 
distributed in both lines, with absorption near Sgr~B2~(M) and (N),
like the SO line (above). The ratio between the isotopologue $^{29}$SiO
and SiO, indicates that the SiO is optically thick. This leads to 
the differences in the integrated distribution, and fitted mean velocity
and velocity width in Table \ref{tab:fitted_peaks}, as the suppression of 
emission in the SiO line centre shifts the mean and increases the fitted width.  
The SiO 1~--~0 line 
distribution coincides with the peak labelled SiO+69-0.06 of this line
in \citet{ma+97}, and has similar absorption at  Sgr~B2~(M) and (N) seen
in the 3-mm SiO 2~--~1 (86.85 GHz) line in that paper.
It is also similar to the distribution of the SiO 2~--~1 line
in \citet{mi07} and \citet{jo+08}, but the latter is at lower 
signal to noise.

We also have data for the  maser lines SiO 1~--~0 v=1 (43.122 GHz) and 
1~--~0 v=2 (42.821 GHz) but did not detect these lines above a 3~$\sigma$ 
level of 0.10 K, although these lines have been detected in Sgr~B2 by, for
example \citet{za+09}. If the v=1 line was a strong as that reported by
\citet{za+09}, corresponding to 0.18 K, it would have been detected here,
but such masers are variable, so it is quite plausible that the strength has 
changed.

\subsection{CS, $^{13}$CS and C$^{34}$S}
\label{subs:cs}

There are three lines of carbon monosulphide CS 1~--~0 (48.99 GHz)
and the isotopologues  $^{13}$CS 1~--~0 (46.25 GHz) and C$^{34}$S
1~--~0 (48.21 GHz) shown in Fig. \ref{fig:SO_etc_int}. Like the SiO line above,
there is widespread emission, with absorption near Sgr~B2~(M) and (N). 
There are optical depth effects which make the integrated emission of the
CS line and the isotopologues  $^{13}$CS 1~--~0 and C$^{34}$S different.
The peaks in the data cubes (Table \ref{tab:fitted_peaks})
are near Sgr~B2~(M) at
peak position $l = 0.661$ deg, $b = -0.041$~deg (mean of the five lines).
The CS line shows absorption, so is fitted in Table \ref{tab:fitted_peaks}
as two gaussians, offset in velocity. 
The peaks of integrated emission (Fig. \ref{fig:SO_etc_int}) are at
different positions, due to the effects of absorption.
These distributions are similar to that 
found by \citet{jo+08} for the 3-mm lines CS 2~--~1 (97.98 GHz)
and the isotopologues  $^{13}$CS 2~--~1 (92.49 GHz) and C$^{34}$S
2~--~1 (96.41 GHz).
The CS 1~--~0 (48.99 GHz) line has also been mapped by \citet{sa+00},
with the Nobeyama 45-m telescope at higher resolution, showing similar
line distribution.

\subsection{CCS}
\label{subs:ccs}

There are two lines of dicarbon monosulphide CCS
shown in Fig. \ref{fig:SO_etc_int}, the 3,2~--~2,1 (33.75 GHz) 
and 4,3~--~3,2 (45.38 GHz) lines. The emission peaks in the north cloud,
with mean peak position $l = 0.688$ deg, $b = -0.014$~deg, 
mean velocity 66~km~s$^{-1}$ and mean velocity width 21~km~s$^{-1}$.
The integrated emission avoids the Sgr~B2~(M) and (N) area, but this may
be due to absorption, as shown in the CCS 45.38 GHz line 
(Fig \ref{fig:SO_etc_int}).
This is different to the distribution of the CCS 7,6~--~6,5 (81.505 GHz) line 
shown by \citet{kusn94}, at higher resolution, which peaks near Sgr~B2~(M) 
and (S).

\subsection{CH$_{3}$CN}
\label{subs:ch3cn}

There is one line of methyl cyanide CH$_{3}$CN, shown in Fig. 
\ref{fig:SO_etc_int}, the 2(0)~--~1(0) group (36.80 GHz). The emission
also peaks in the north cloud, and shows the ridge in the integrated emission
to the north-west of Sgr~B2~(N) and (M), similar to that in the CH$_{3}$CN
5~--~4 (91.99 GHz) and 6~--~5 (110.38 GHz) lines in \citet{jo+08} and the 
5~--~4 line in \citet*{demawi97}. There may also be some absorption in the 
2(0)~--~1(0) line here, around Sgr~B2~(N) and (M), not just a deficit of 
emission.

\subsection{OCS}
\label{subs:ocs}

\begin{figure*}
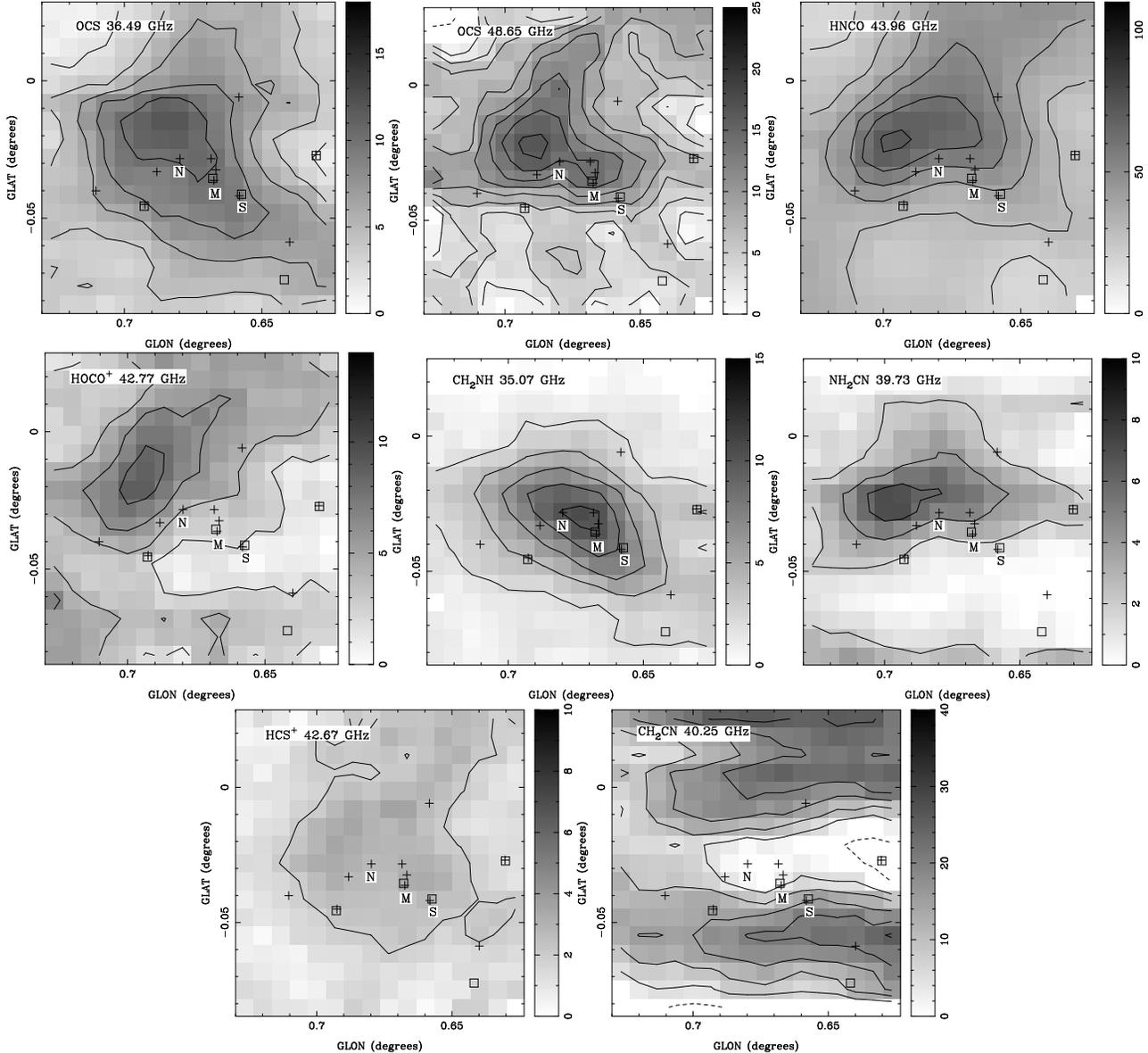

\includegraphics[width = 5.2 cm,angle=-90]{SgrB2_2009_7mm_OCS_B_int3.ps}
\includegraphics[width = 5.2 cm,angle=-90]{SgrB2_2009_7mm_OCS_A_int3.ps}
\includegraphics[width = 5.2 cm,angle=-90]{SgrB2_2009_7mm_HNCO_int3.ps}
\includegraphics[width = 5.2 cm,angle=-90]{SgrB2_2009_7mm_HOCO+_int3.ps}
\includegraphics[width = 5.2 cm,angle=-90]{SgrB2_2009_7mm_CH2NH_int3.ps}
\includegraphics[width = 5.2 cm,angle=-90]{SgrB2_2009_7mm_NH2CN_int3.ps}
\includegraphics[width = 5.2 cm,angle=-90]{SgrB2_2009_7mm_HCS+_int3.ps}
\includegraphics[width = 5.2 cm,angle=-90]{SgrB2_2009_7mm_CH2CN_int3.ps}
\caption{The integrated emission images for two lines of OCS, and one line each
of HNCO, HOCO$^{+}$, CH$_{2}$NH, NH$_{2}$CN, HCS$^{+}$ and CH$_{2}$CN.}
\label{fig:misc_int}
\end{figure*}

There are two lines of carbonyl sulphide OCS
shown in Fig. \ref{fig:misc_int}, the 3~--~2 (36.49 GHz) and 4~--~3 
(48.65 GHz) lines. The emission peaks in the north cloud,
with mean peak position $l = 0.684$ deg, $b = -0.028$~deg, 
mean velocity 63~km~s$^{-1}$ and mean velocity width 22~km~s$^{-1}$,
although there are some differences in distribution between the two
transitions. The distribution is similar to the  7~--~6 (85.14 GHz),
8~--~7 (97.30 GHz) and 9~--~8 (109.46 GHz) lines \citep{jo+08},
with the ridge near Sgr~B2~(N) and (M).

\subsection{HNCO and HOCO$^{+}$}
\label{subs:hnco+hoco+}

There is one line of isocyanic acid HNCO the 2(0,2)~--~1(0,1) group (43.96 GHz)
and one of protonated CO$_{2}$ HOCO$^{+}$ 2(0,2)~--~1(0,1) (42.77 GHz)
shown in Fig. \ref{fig:misc_int}. These lines peak at the north cloud,
so the two molecules are discussed here together. 

The HNCO 43.96 GHz distribution is similar to that of the 4(0,4)~--~3(0,3)
(87.93 GHz) and 5(0,5)~--~4(0,4) (109.91 GHz) lines in \citet{jo+08}
and \citet{mi+98}. Observations of the 4(0,4)~--~3(0,3) line over a wider area
(albeit at lower resolution) are shown in \citet{miir06} along with discussion
of this north cloud (called the 2'N HNCO peak by them). There are also
observations of the 1(0,1)~--~0(0,0) (21.98 GHz) line in \citet{wi+96},
with similar distribution.

The HOCO$^{+}$ 42.77 GHz distribution similar to that of the 4(0,4)~--~3(0,3)
(85.53 GHz) and 5(0,5)~--~4(0,4) (106.91 GHz) lines in \citet{jo+08},
\citet{mi+98} and \citet*{miirzi88} and similar to the HNCO lines,
highlighting the chemical distinctness of this north cloud \citep{wi+96}
from the Sgr~B2~(M) and (N) area.

\subsection{CH$_{2}$NH, NH$_{2}$CN, HCS$^+$ and  CH$_{2}$CN}
\label{subs:misc}

Finally, there is one line of each of the methylenimine CH$_{2}$NH 
3(0,3)~--~2(1,2) group (35.07 GHz), cyanamide NH$_{2}$CN 2(1,2)~--~1(1,1),v=0 
(39.73 GHz), thioformyl HCS$^+$ 1~--~0 (42.67 GHz)
and the cyanomethyl radical CH$_{2}$CN 2~--~1 group (40.25 GHz)
shown in Fig. \ref{fig:misc_int}.

The CH$_{2}$NH 35.07 GHz line distribution peaks near Sgr~B2~(M), 
and is similar to the distribution of the 4(0,4)~--~3(1,3) (105.79 GHz) line
in \citet{jo+08}. The latter being at higher resolution resolves the structure
better, showing the peak at Sgr~B2~(N) whereas the fitted peak in the 35.07
GHz line is near  Sgr~B2~(M). There are likely to be excitation differences 
between the two cores.

The NH$_{2}$CN 39.73 GHz line peaks at the north cloud, with the ridge to the 
north-west of Sgr~B2~(M), with a distribution similar to that of the 
5(1,4)~--~4(1,3) (100.63 GHz) line \citep{jo+08}, although the image there
is noisy and affected by scanning stripes.

The HCS$^+$ 42.67 GHz line is quite weak, but has an extended distribution.
Not much is known on the distribution of this ion in Sgr B2, but it is plausible
that the distribution is similar to that of CS (e.g. Fig. \ref{fig:SO_etc_int},
subsection \ref{subs:cs}).

The distribution of integrated emission of the CH$_{2}$CN 40.25 GHz line
is extended, with likely absorption near Sgr~B2~(N). 
There are some stripes in the integrated emission in the longitude scanning
direction.
The distribution differs from
that of CH$_{3}$CN (Fig. \ref{fig:SO_etc_int},
subsection \ref{subs:ch3cn}) which may indicate that the two molecules have 
quite different formation routes and destruction paths \citep{ir+88, tu+90}
despite their apparent similarity. 

\subsection{Summary of peak positions}
\label{subs:peaks}

\begin{figure}
\includegraphics[width = 8.5 cm,angle=-90]{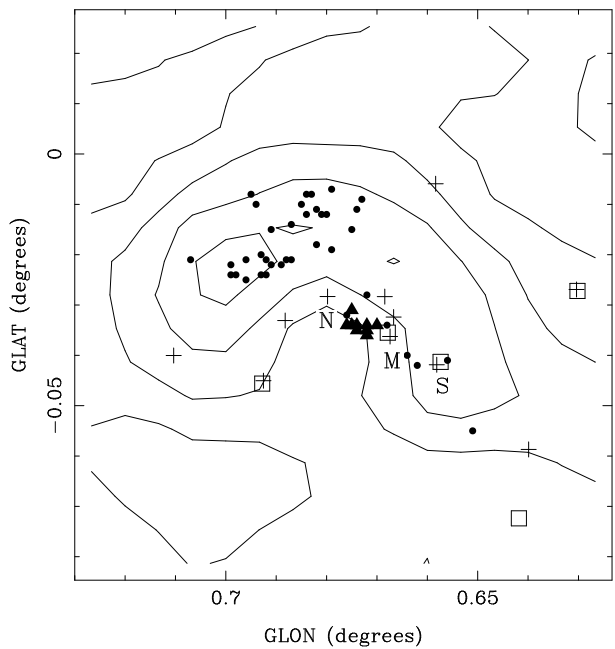} 
\caption{The positions of the fitted 7-mm peaks, with filled circles for the 
molecular peaks and filled triangles for the H radio recombination lines.
The contours are from the CH$_{3}$CN line. The crosses are radio peaks and 
open squares infrared peaks, as in other images.}
\label{fig:fit_posns}
\end{figure}

The peak positions of the 7-mm lines in the data cubes tabulated in Table 
\ref{tab:fitted_peaks} are plotted in Fig. \ref{fig:fit_posns}.
The hydrogen recombination lines (filled triangles in Fig. \ref{fig:fit_posns})
peak between Sgr B2 (N) and (M), as discussed in subsection \ref{subs:rrls}
and shown in Fig. \ref{fig:RRL_7mm_overlay}. The molecular lines mostly peak 
at the north cloud. 

We attribute the molecular line distributions in this
7-mm band to a combination of the effect of emission, from an extended
cloud centred to the east of Sgr B2 (N) and (M), and absorption
due to the free-free continuum at Sgr B2 (N) and (M). We consider this further
in the Discussion (Section \ref{sec:discuss}) and in the next section 
(Section \ref{sec:pca}) where we analyse the spatial distributions in a relatively
objective way. 

\section{Principal Component Analysis}
\label{sec:pca}

The integrated emission images of the different lines shown in
Figs. \ref{fig:RRLs_int} and \ref{fig:HC3N+HC5N_int} to \ref{fig:misc_int} show 
similarities and 
differences. One useful technique to identify and quantify the features of 
large data sets is Principal Component Analysis (PCA), see e.g \citet{hesc97}. 
This describes
the multi-dimensional data set by linear combinations of the data that describe
the largest variance (the most significant common feature) and successively
smaller variances (the next most significant features). 

In the context here,
we can use PCA to describe the large number of images, with a smaller set of
images which contain the most significant features. This has been used by 
\citet{un+97} for the OMC-1 ridge, and more recently by us for the G333 
molecular cloud \citep{lo+09} and the Sgr B2 area with 3-mm molecular lines
\citep{jobulo08}.

We have implemented the PCA processing in a {\sc python} script, with the 
PCA module\footnote{http://folk.uio.no/henninri/pca\_module/}, and 
pyFITS\footnote{http://www.stsci.edu/resources/software\_hardware/pyfits/} 
to read and write the FITS images.  

\begin{figure*}
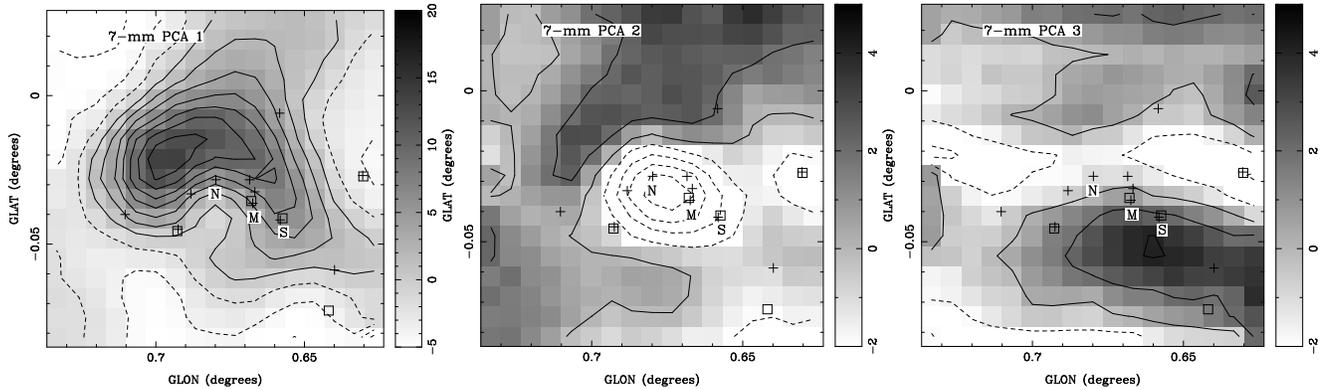

\includegraphics[width = 5.1 cm,angle=-90]{SgrB2_7mm_PCA_1_im3.ps}
\includegraphics[width = 5.1 cm,angle=-90]{SgrB2_7mm_PCA_2_im3.ps}
\includegraphics[width = 5.1 cm,angle=-90]{SgrB2_7mm_PCA_3_im3.ps}
\caption{The first three images from the Principal Component Analysis (PCA)
for the 7-mm lines, describing 56, 14 and 9~percent of the data variance 
respectively. The first image highlights emission from the north cloud 
and ridge, the second image highlights absorption (or emission) from Sgr B2 (N) 
and (M) and the third image highlights differences in emission to the south.}
\label{fig:PCA_im_7mm}
\end{figure*}

The images of the three most significant components of the 7-mm integrated 
emission are shown in Fig. \ref{fig:PCA_im_7mm}. The first four components 
describe respectively 56, 14, 9 and 4~percent of the variance in the data.
These components are statistical descriptions of the integrated line images,
not necessarily physical components of Sgr B2. However, they do highlight
the physical features in a useful way.

The first component (Fig. \ref{fig:PCA_im_7mm}) highlights emission from the 
north cloud and ridge. The second component highlights absorption (or emission) 
from Sgr B2 (N) and (M), and also some differences in the north cloud.
The third component highlights differences in emission to the south. The
fourth component (not shown here in Fig. \ref{fig:PCA_im_7mm}) highlights 
differences in emission from Sgr B2 (N).

\begin{figure*}
\hspace{-7mm}
\includegraphics[width = 6.4 cm]{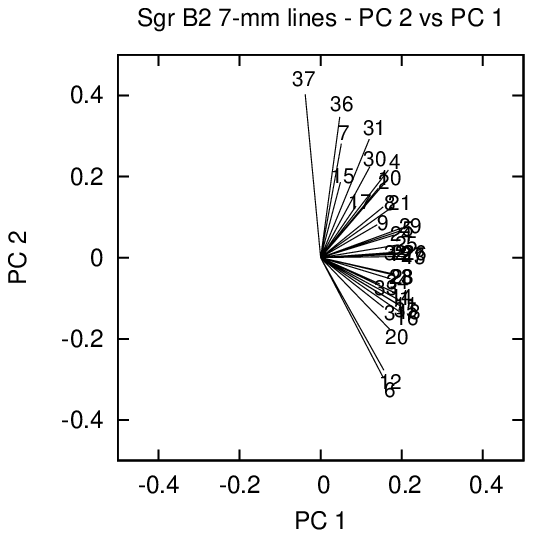}
\hspace{-7mm}
\includegraphics[width = 6.4 cm]{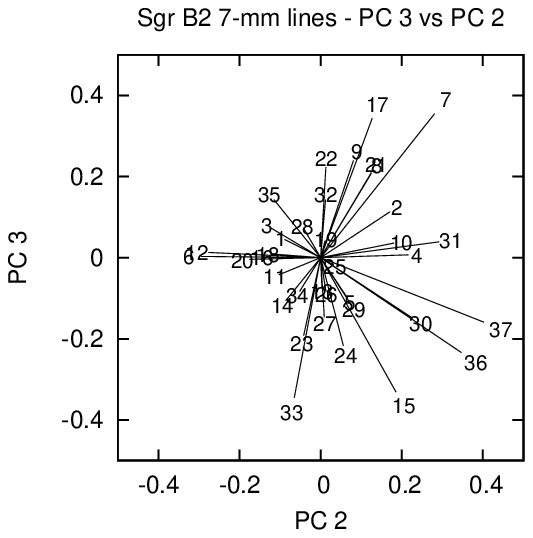}
\hspace{-7mm}
\includegraphics[width = 6.4 cm]{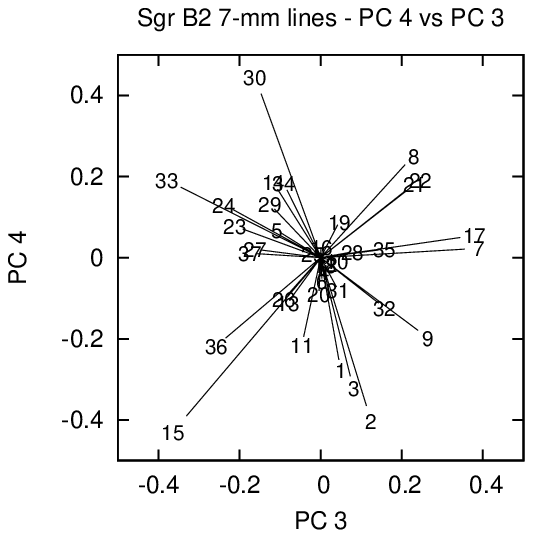}
\caption{The component vectors of the integrated line images in the 
decomposition of the data into the PCA images, shown with successive pairs of 
PCA images. The vectors are labelled with numbers, rather than molecule names, 
for clarity of presentation, with the molecules listed in alphabetical order:
1, $^{13}$CS; 2, $^{29}$SiO;3, C$^{34}$S; 4 and 5, two lines of CCS;
6, CH$_{2}$NH; 7 to 10, four lines of CH$_{3}$CHO; 11, CH$_{3}$CN;
12 to 14, three lines of CH$_{3}$OH; 15, CS; 16 and 17, two lines of 
H$^{13}$CCCN; 18 to 24, seven lines of HC$_{5}$N; 25, HCC$^{13}$CN; 
26 and 27, two lines of HC$_{3}$N; 28, HCS$^{+}$; 29, HNCO; 30, HOCO$^{+}$; 
31 and 32, two lines of NH$_{2}$CHO; 33, NH$_{2}$CN; 34 and 35, two lines of 
OCS; 36, SiO, and; 37, SO.}
\label{fig:PCA_vect}
\end{figure*}

The relations between the molecular integrated emission images, and the PCA
images are shown in Fig. \ref{fig:PCA_vect} as vectors of the projection
of the data images on the axes of the PCA images. The molecules are labelled
with numbers, for clarity, in these plots. All of the data images are
positively correlated with the first PCA image, except for SO (37).
The greatest positive correlation with the second PCA component are for
SO (37) and SiO (36), which have strong absorption at  Sgr B2 (N) and (M).
The greatest negative correlation with the second PCA component are for
CH$_{2}$NH (6) and the CH$_{3}$OH 44.07 GHz line (12, with some maser emission), 
which have emission at Sgr B2 (N) and (M).
The greatest positive correlation with the third PCA component are for
the CH$_{3}$CHO 38.512 GHz line (7) and the H$^{13}$CCCN 35.27 GHz line (17),
and greatest negative correlation with the third PCA component are for
NH$_{2}$CN (33) and CS (15), highlighting more or less emission to the south,
relative to the other lines.

\begin{figure*}
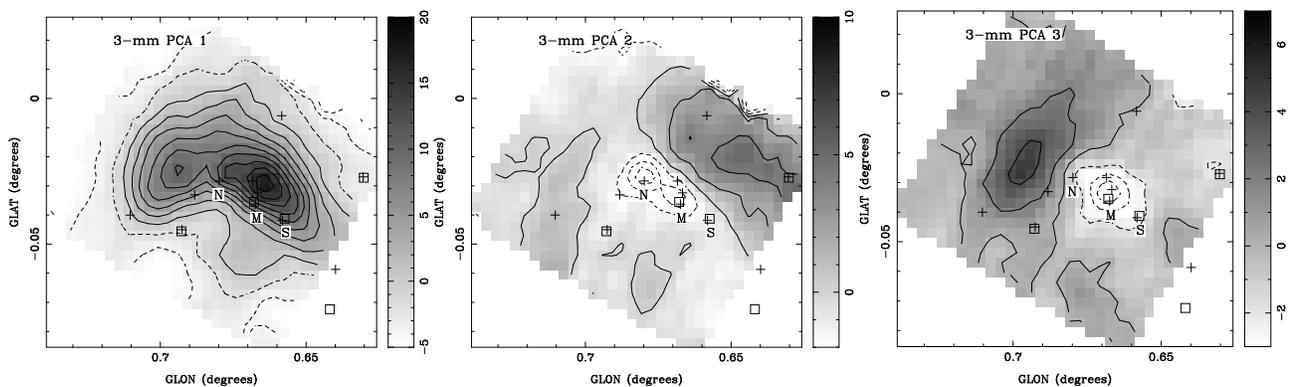

\includegraphics[width = 5.0 cm,angle=-90]{SgrB2_3mm_PCA_1_gal_im3.ps}
\includegraphics[width = 5.0 cm,angle=-90]{SgrB2_3mm_PCA_2_gal_im3.ps}
\includegraphics[width = 5.0 cm,angle=-90]{SgrB2_3mm_PCA_3_gal_im3.ps}
\caption{The first three images from the Principal Component Analysis (PCA)
for the 3-mm lines (lower) from \citet{jo+08}. These 3-mm images are 
similar to those in \citet{jobulo08} rotated into 
Galactic coordinates for easier comparison to the 7-mm images. 
The first image highlights emission from the north cloud 
and ridge. The second image highlights absorption (or emission) from Sgr B2 (N) 
and (M) and emission from the south-west. The third image highlights 
the north cloud and differences in emission from Sgr B2 (M).}
\label{fig:PCA_im_3mm}
\end{figure*}

We used the integrated emission images for 37 molecules here in the PCA, 
excluding CH$_{2}$CN  (as it is probably affected by the scanning ripples),
and not including the radio recombination lines (RRLs), as they are due to a 
different  
physical mechanism. However, we have also calculated the PCA for the 46 integrated
emission images, including the RRLs, and find very similar
results. The first four components then 
describe respectively 49, 24, 8 and 3~percent of the variance in the data,
with the PCA images quantitatively very similar, with the major difference being
the second component image having a stronger peak at Sgr B2 (N) and (M),
describing the RRLs, and positive rather than negative. However, the sign of
the component images is not physically significant.

The PCA images from these Sgr B2 7-mm lines can be compared to those found
for the Sgr B2 3-mm lines from \citet{jo+08} in Fig. \ref{fig:PCA_im_3mm} 
\citep{jobulo08}. The 3-mm PCA images are at higher resolution.

The first two PCA component images are similar: the first
highlighting the ridge and north cloud, and the second highlighting absorption 
at Sgr B2 (N) and (M). The second PCA component at 3-mm also includes emission
to the south-west, as some of the strongest lines at 3-mm, such as HCN, 
HCO$^{+}$ and HNC are optically thick with both absorption at Sgr B2 (N) and 
(M) and the south-west area relativly stronger (as the the main emission is 
relatively weaker due to the high optical depth). The third 3-mm PCA
component highlights the north cloud, and Sgr B2 (M). As PCA is a statistical 
tool, it is not expected that the higher PCA images, of lower statistical 
weight, should be the same from the independent 7-mm and 3-mm data. 
However, the PCA images from the 3-mm and 7-mm data do show quite similar
spatial features, albeit distributed differently amongst the PCA images:
ridge, north cloud, Sgr B2 (N) and Sgr B2 (M).

\section{Discussion}
\label{sec:discuss}

We can consider the distribution of the spectral lines in terms of physical
components of the Sgr~B2 complex: the hot cores Sgr B2 (N), (M) and (S),
and the extended envelope. These are related to the Principal Components
discussed in the last section, but are not the same decomposition of structure
into features. In \citet{jo+08} we identified seven features. Of these,
the extended ridge and north cloud are 
related to the extended envelope of Sgr~B2
\citep{jo+08} which is traced by the dust seen at sub-mm wavelengths
(Fig. \ref{fig:radio_submm_ir}). The hot cores Sgr B2 (N), (M) and (S)
are denser and warmer, and so have different chemistry, so they are prominent
in particular molecules, as shown in Fig. \ref{fig:misc_int} by CH$_{2}$NH.
In the 7-mm band data here, there is also often absorption in the area
around Sgr B2 (N) and (M) related to the radiative transfer of the continuum
free-free emission through the molecular gas.

To better describe the quantitative properties of these physical components of
Sgr~B2, we have considered in the Appendix continuum data (imaging and 
photometry) from the radio, through sub-mm to mid-IR wavelenths, from the 
literature,
plus some of our own (as yet unpublished) radio and sub-mm data, and data from
public archives. We have fitted the spectral energy distribution (SED) of
the Sgr~B2 (N), (M) and (S) cores, and the extended envelope, using the imaging
to set the angular size of the components, to derive parameters such as dust
temperature.

The integrated line emission $\int T_{B} dv$ plotted in Figs. 
\ref{fig:HC3N+HC5N_int} 
to \ref{fig:misc_int} is related the the column density of the molecule
in the upper state of the transition $N_{u}$ by 
\begin{equation} 
N_{u} = (8 \pi \nu^2 k/h c^3 A_{ul}) \int T_{B} dv 
\end{equation}
where $A_{ul}$ is the Einstein coefficient, and assuming optically thin 
emission (optical depth $\tau \ll 1$). This is related to the total column 
density of the molecule $N$, assuming local thermodynamic
equilibrium (LTE) by 
\begin{equation}
N = N_{u} (Q_{T}/g_{u}) \exp(E_{u}/kT_{ex}) 
\end{equation}
where  $g_{u}$ is the statistical weight of the upper level, $Q_{T}$ is the 
partition function at excitation temperature $T_{ex}$ and $E_{u}$ is the
energy of the upper level. This analysis also assumes the line emission
fills the beam (filling factor unity) and that the line brightness
temperature $T_{B}$ is much greater than the continuum brightness
temperature $T_{C}$.

In this simple case, then, the integrated line 
emission $\int T_{B} dv$ traces the molecule column density $N$, and we can use
this to trace chemical variations in the Sgr~B2 complex. However, the 
simplifying assumptions are often not valid, so the relation is more
complicated.

Firstly, some transitions are not in LTE, as shown by the differences
between the three methanol (CH$_{3}$OH) lines in Fig. \ref{fig:CH3OH_etc_int},
which include maser transitions. Secondly, even if the lines are in {\it local}
thermodynamic equilibrium, there are large gradients in the kinetic temperature
$T_{K}$ and radiation environment (corresponding to radiation temperature 
$T_{r}$) giving gradients in excitation temperature $T_{ex}$. So transitions
from higher energy states will preferentially highlight higher temperature 
regions. Thirdly, the lines are often not optically thin, so that the 
optically thick emission is not simply proportional to the column density 
$N_{u}$, in the sense that the integrated intensity saturates for high optical 
depth in regions of high column density. Fourthly, we have, so far, 
ignored the full radiative transfer model in which background continuum
radiation can be absorbed by the molecules in the cloud, as well as the 
molecules
emitting. With a background continuum source of brightness temperature $T_{C}$, 
we obtain 
\begin{equation}
\begin{tabular}{cll}
$ T_{B} $ & = & $ T_{ex}[1 - \exp(-\tau)] + T_{C}\exp(-\tau) $ \\
          & = & $ T_{ex} + (T_{C} - T_{ex})\exp(-\tau) $ \\
  & $\approx$ & $ T_{C} + (T_{ex} - T_{C})\tau$ for $\tau \ll 1$. \\
\end{tabular}
\end{equation}
Typically, the spectral line data (as in this paper) are baselevel subtracted,
to be the line only, but the effect of the background continuum is to reduce 
the line intensity, and for $T_{C} > T_{ex}$ the line becomes negative in
intensity, i.e. is seen in absorption. 

Most of the lines presented here are dominated by emission from the extended,
cooler ($T_{dust} \sim 20$~K), envelope, rather than the compact hot cores. The 
scale of this envelope
is around 2 arcmin, but elongated north-south in equatorial
coordinates (as shown in the mid-IR, Fig. \ref{fig:radio_submm_ir}). This 
envelope is centred to the east of the line of hot cores Sgr~B2 (N), (M) and (S),
with the north cloud to the north of the hot cores. In Galactic coordinates,
this corresponds to a ridge, wrapping around the hot cores, as shown in the 
first principal components of the 7-mm (Fig. \ref{fig:PCA_im_7mm}) and 3-mm 
lines (Fig. \ref{fig:PCA_im_3mm}). There are also chemical differences,
for example with HNCO and HOCO$^{+}$ preferentially found 
(Fig. \ref{fig:misc_int}) in the north cloud. The effective spatial
scale of the stronger, optically thick lines, such as SO, SiO and CS 
(Fig. \ref{fig:SO_etc_int}) is larger, with the emission filling the whole 
$6 \times 6$~arcmin$^{2}$ area imaged, as the central peak emission saturates
(and there is absorption). The optically thin isotopologues ($^{29}$SiO, 
$^{13}$CS and C$^{34}$S, Fig. \ref{fig:SO_etc_int}) are better tracers of the
extent of the envelope.

The recombination lines (Fig. \ref{fig:RRLs_int}) show the distribution
of the 7-mm free-free radio continuum, assuming that there is not a very
large spatial variation in the line to continuum ratio (so that the line
emission traces the continuum distribution). The deconvolved
angular size (from these recombination lines) is $2.2 \times 0.9$ arcmin$^{2}$,
and the fit to the radio flux densities (Fig. \ref{fig:SEDs} and 
Table \ref{tab:SED_fits}) shows that around half the flux comes from the 
extended envelope and half from the compact hot cores. The spatial distribution
along the line of sight of the Sgr~B2 components is not well constrained, 
but the spectral energy distribution (SED) fitting (Appendix) does indicate
that Sgr~B2 (N) is behind at least part of the dust envelope. Many of the 7-mm
molecular lines in Figs. \ref{fig:HC3N+HC5N_int} to \ref{fig:misc_int} show 
absorption centred near Sgr~B2 (N) and (M) which matches well the position
and scale of the 7-mm free-free continuum (as probed by the recombination lines
in Fig. \ref{fig:RRLs_int}). 

With a flat spectral index 
($\alpha =-0.17 \pm 0.07$ for the total free-free) in the optically thin regime,
the continuum brightness temperature falls steeply with frequency
$T_{C} \propto \nu^{\alpha - 2}$. The conditions for absorption of the continuum
radiation, by the intervening cool envelope ($T_{C}$ significant compared 
to $T_{ex}$), are commonly met in these 7-mm 
data, as highlighted in the second principal component of Fig. 
\ref{fig:PCA_im_7mm}. The molecules with the most significant absorption
are SO, SiO, $^{29}$SiO, CS, CH$_{3}$CHO, NH$_{2}$CHO and CCS (although for the 
last three molecules this is less clear in some of the lower signal to noise
lines).

Similar absorption occurs for the 3-mm data of 
\citet{jo+08} (seen in the second principal component of Fig. 
\ref{fig:PCA_im_3mm}). At these higher frequencies, the absorption
at the hot cores
Sgr~B2 (N) and (M) dominates over the absorption due to the extended envelope.
The resolution is higher, the cores contribute
a larger fraction of the free-free continuum, and have a higher brightness 
temperature. 

In both the 7-mm and 3-mm molecular lines, there are also
lines that peak at the Sgr~B2 (N) and (M) hot cores, due to their
different chemical conditions and higher temperature. In the 7-mm data here,
the most prominent `hot core' lines are from CH$_{2}$NH, and the 44.07 GHz
transition of CH$_{3}$OH which includes maser emission \citep{meme97}.

The chemistry of molecular clouds is complex \citep{heva09}, particularly
those associated with star-formation, like Sgr B2. Broadly, the chemistry
of regions that are heated by star formation to become hot cores can
be described in three stages: the cold, warm up and hot core stages.

In the cold stage, there are gas-phase reactions plus grain-surface reactions,
with a lot of the complex chemistry occuring between molecules sticking
on the ice-grains. In the warm up stage, as the material forming the protostar
collapes, there is heating (from 10 K to a few hundred K), and the 
grain mantles sublime, releasing different 
molecules at different temperatures and hence times. There are further gas-phase
and grain-surface reactions occuring with modified chemical abundances
and hotter temperatures. In the last hot core stage, with temperatures 
100 to 300 K, the ices have sublimed, so there is only gas-phase chemistry.

The overall chemical evolution is complicated, and time dependent, so detailed 
models typically consider a particular phase. A recent gas-grain warm up model,
of relevance to Sgr B2 (N) is \citet{gawehe08}. 

The observed spectral-line data can be compared to such models, and constrain
models, but this is not straight-forward \citep{heva09}. For a particular 
position,
or an unresolved source, the observed antenna temperature can be converted
to a source brightness temperature, using the source size or filling factor.
This can be converted to a molecule column density (or abundance relative 
to e.g. CO) using the excitation temperature $T_{ex}$ and optical depth $\tau$
(preferably using constraints from other lines). A better way to analyse
the molecular line data is to construct a physical model of the source
\citep[see][Figure 4]{heva09} and calculate the full non-LTE statistical 
equilibrium excitation of the molecule through the physical structure.

For the Sgr B2 complex data presented here, with the hot cores, and cooler 
envelope, we do not attempt such a detailed radiative transfer model, but
simply point out the broad features required in such a model, namely
the physical components (see Appendix) and the effects of optical depth and 
absorption. We show by the line distribution images that there are
significant spatial differences in the chemical distribution, that can
be interpreted in terms of the cool, warm up and hot core `stages' of the 
chemical models, with different parts of the Sgr B2 complex having different
heating conditions and history.


\section{Summary}
\label{sec:summary}

We have undertaken a line imaging survey of the Sgr~B2 region from 30 to 50 GHz 
(the 7-mm band) with the 22-m single-dish Mopra telescope.  This complements 
the Mopra 3-mm line imaging of Sgr~B2 presented in \citet{jo+08}. Integrated 
emission images of 47 lines are presented: 38 molecular lines and 9 radio 
recombination lines.

The distributions of individual lines are discussed, and we have studied
the similarities and differences between the lines with principal
component analysis (PCA). The major features from the PCA are extended emission
from the ridge and north cloud to the north and east of the Sgr~B2 (N) and (M)
hot cores, and absorption near  Sgr~B2 (N) and (M). These statistical features
are interpreted in terms of the physical components of the Sgr~B2 complex 
(discussed in the appendix) of the extended low temperature envelope,
and compact hot cores Sgr~B2 (N), (M) and (S). The 7-mm free-free emission
in the area of Sgr~B2 (N) and (M) is absorbed in this line-of-sight
by the envelope.

\section*{Acknowledgments}

The Mopra telescope is funded by the Commonwealth of Australia as a National
Facility managed by CSIRO as part of the Australia Telescope.
The UNSW-MOPS Digital Filter Bank used
for the observations with the Mopra telescope was provided with
support from the Australian Research Council (ARC), together with the
University of New South Wales, University of Sydney and Monash
University. We also acknowledge ARC support through Discovery Project
DP0879202. PAJ acknowledges partial support 
from Centro de Astrof\'\i sica FONDAP 15010003 and the Gemini-CONICYT Fund.
We thank John B. Whiteoak for the 1.2-mm SIMBA data,
and an anonymous referee for useful comments that improved
the presentation of the paper.


\appendix

\section{The components of Sgr~B2}

\begin{figure*}
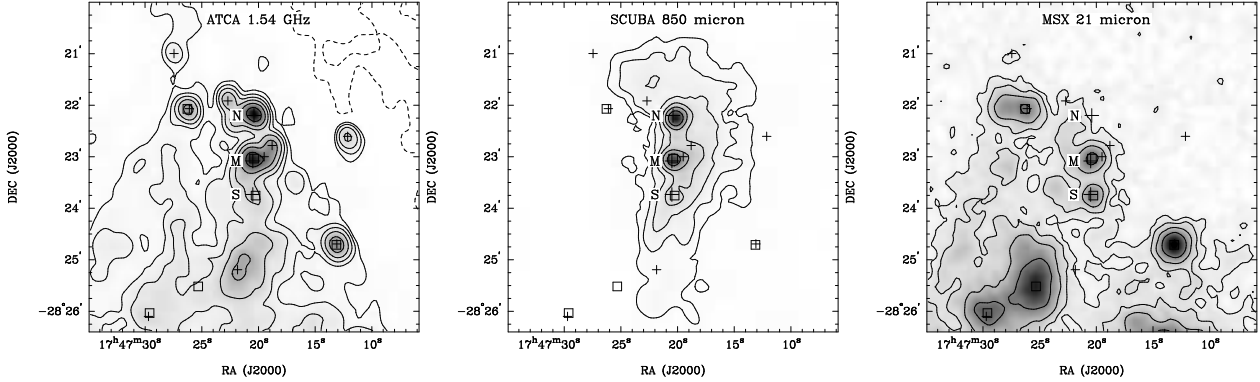

\includegraphics[width = 5.0 cm,angle=-90]{ATCA_19cm_SgrB2_p2.ps}
\includegraphics[width = 5.0 cm,angle=-90]{SCUBA_850_SgrB2_p2.ps}
\includegraphics[width = 5.0 cm,angle=-90]{MSX_21_SgrB2_p2.ps}
\caption{The continuum images of the Sgr~B2 complex in radio (ATCA 1.54 GHz),
sub-mm (SCUBA 850 $\mu$m) and mid-infrared (MSX 21 $\mu$m), showing the 
Sgr B2 (N), (M) and (N) cores. The radio and mid-infrared trace star formation,
with some extra compact features, and extended emission to the south-east.
The sub-mm traces warm and cold dust, including diffuse emission around the 
cores. Note that these images are in equatorial coordinates, rather than
Galactic coordinates used in the main body of this paper, but the crosses for
radio peaks and open squares for infrared peaks are used as fiducial marks.}
\label{fig:radio_submm_ir}
\end{figure*}

The Sgr~B2 complex is very complicated, with many physical components
distinguishable at high resolution, see for example \citet{meme97} and 
\citet{ga+95}.
However, for many observations with resolution of order tens of arcseconds
(Fig. \ref{fig:radio_submm_ir})
it is common, and physically useful, to consider the three dense cores
Sgr~B2 (N), (M) and (S), and distinguish them from the larger and less dense 
envelope. 

The three dense cores are prominent in both the radio free-free and sub-mm dust
continuum, as shown in Fig. \ref{fig:radio_submm_ir}, and the (M) and (S) cores
are prominent in the mid-infrared. The sub-mm dust distribution also shows
the extended component, centred slightly to the west of the line of dense 
cores, with an extension to the north. As we pointed out in \citet{jo+08},
many of the 3-mm molecular lines follow the ridge of this extended dust 
component, while other lines are associated with the dense cores.
We find similar distributions here for the 7-mm molecular lines, so for more
detailed quantitative discussion of these 7-mm and 3-mm line data, we have 
undertaken an analysis of the Sgr B2 continuum data.

The structure of the Sgr B2 complex has been analysed by \citet{go+93}
using the spectral energy distribution (SED) of the total complex
and components FIR1, FIR2 and FIR3
in their nomenclature, corresponding to (N), (M) and (S).

The peak in the SED due to dust is modelled as a Black Body modified
by the optical depth $\tau_{\nu}$ of the dust, so the flux density $S_{\nu}$
divided by the source solid angle $\Omega$ is 
$ S_{\nu}/\Omega = B_{\nu}(T) (1 - \exp(-\tau_{\nu})) $
where $B_{\nu}(T)$ is the Planck function \citep*{wirohu09}.
The dust optical depth is assumed to be proportional to the gas column density,
with some empirical scaling factor. The model from \citet*{mezywi90} is used
with 
$ \tau_{\nu} = (N_{H}/ 1.4 \times 10^{20} {\rm cm}^{-2}) 
(\lambda/\mu {\rm m})^{-2} b (Z/Z_{\odot}) $,
where $N_{H}$ is the total hydrogen column density,
$(Z/Z_{\odot})$ is the metallicity, assumed 2 for the Galactic Centre
and $b$ is a dust parameter, with $b = 3.4$ for hydrogen number density
$n_{H} \geq 10^{6}$ cm$^{-2}$ and  $b = 1.9$ for $n_{H} \leq 10^{6}$ cm$^{-2}$.

The observed flux densities in the dust SED can be fitted with 3 parameters;
temperature $T$, solid angle $\Omega$ and the wavelength $\lambda_{\tau = 1}$
at which the optical depth is unity, related by the above equation to the
column density $N_{H}$.
(In this model the power-law index of the dust optical depth is $\beta = 2$:
other models can have different power law slopes $\beta$, but we leave $\beta$
fixed here to reduce the number of free parameters in the fit.)

The SED model assumed is simple, with constant temperature $T$, over a uniform 
dust density distribution of solid angle $\Omega$, but this is a common
assumption in fitting sub-mm dust SEDs.
It also has the advantage
of a simple analytic form that can be fitted. We recognise that the dust
distribution and temperature are not uniform (as shown in higher resolution
studies) but argue that the characteristic temperature, density and other
parameters derived (Table \ref{tab:SED_fits}) are useful for quantitative
analysis.

\begin{figure*}
\includegraphics[width = 6.5 cm]{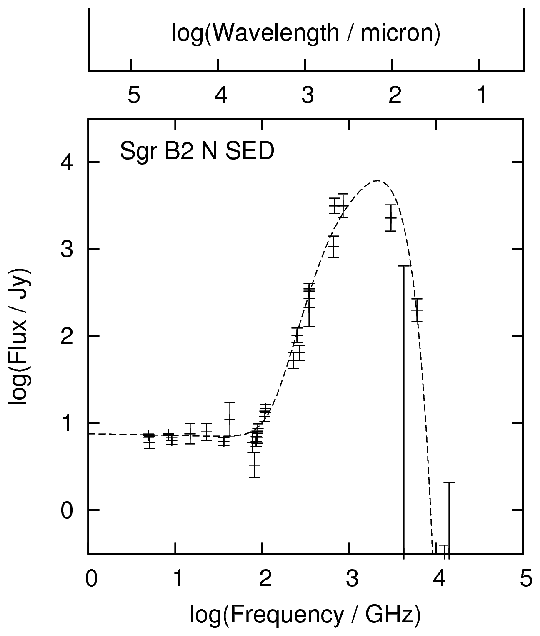}
\includegraphics[width = 6.5 cm]{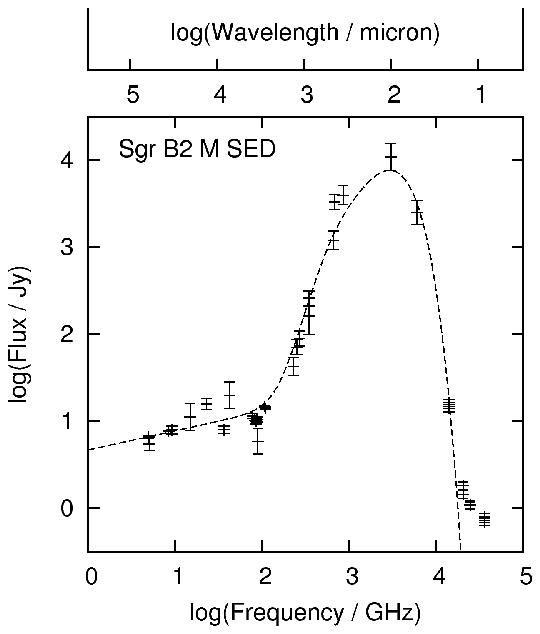}
\includegraphics[width = 6.5 cm]{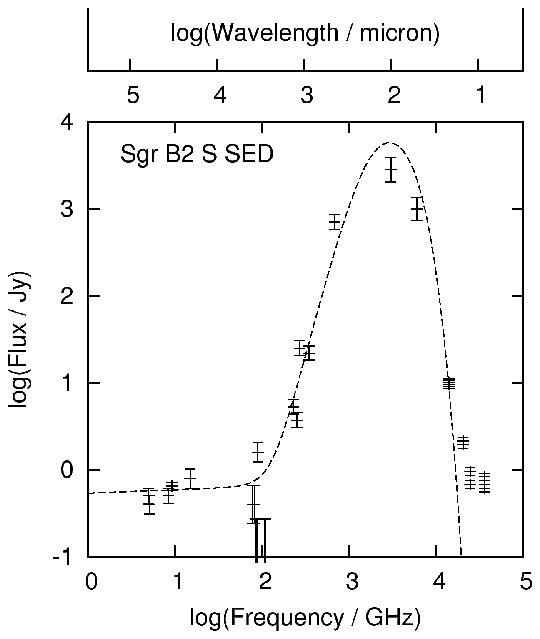}
\includegraphics[width = 6.5 cm]{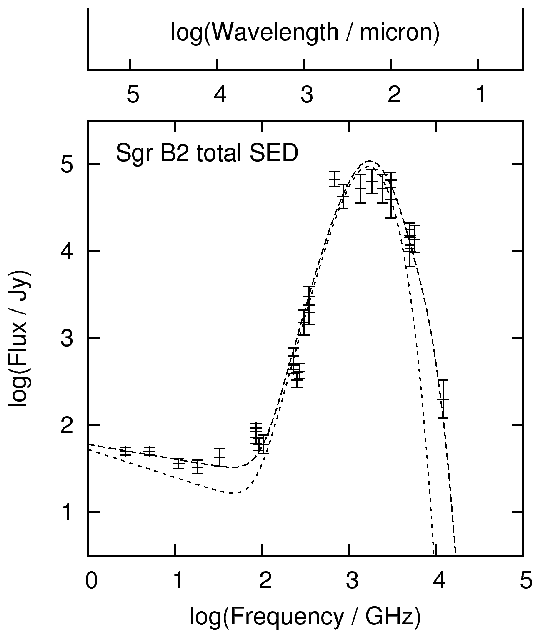}
\caption{The spectral energy distribution (SED) of the Sgr B2 components
(N), (M), (S) and total, from the radio to the infrared, following 
\citet{go+93}, with some additional data. The dashed lines are our schematic
fits to the SED, but we note that there are problems using fluxes from
the literature with a wide range of spatial resolutions so that the $\chi^{2}$
value of the model fits are poor. The plot for the total flux has a long
dashed line for the total flux points, and a shorter dashed line 
for the extended envelope which contains most of the flux. The plot for
Sgr B2 (N) includes 3-sigma upper limits, from Spitzer MIPS at 70 and 24 micron,
and MSX at 21 micron.}
\label{fig:SEDs}
\end{figure*}

We plot in Fig. \ref{fig:SEDs} the SED of the Sgr B2 components, using the data
from \citet{go+93} plus some additional data, from more recent observations.

We include fluxes from \citet{kumesn96} for the compact components in the 3-cm
and 3-mm bands (combining their N and N$'$ into the one source).
We also include mid-infrared fluxes from the MSX 
catalogue Baby Gator 
interface\footnote{http://irsa.ipac.caltech.edu/applications/BabyGator/}.

We have fitted data from our own observations with the ATCA at 1.54 GHz 
(Fig. \ref{fig:radio_submm_ir}) and 36 GHz (Fig. \ref{fig:RRL_7mm_overlay})
and with SIMBA on SEST at 1.2 mm. The fitting was done with multiple
gaussians using the task {\it imfit}
in the {\sc MIRIAD} package, to obtain angular size information as well as 
flux densities. 

We have also fitted data obtained as FITS images from several 
public databases. 
In the radio we obtained VLA images at 20 cm, 3.6 cm, 9 GHz (3.3 cm)
and BIMA data at 3 mm from the Astronomy Digital Image 
Library (ADIL)\footnote{http://imagelib.ncsa.uiuc.edu/imagelib.html.
Note that some of the archived radio images from ADIL were not suitable
for measuring flux densities, as the images were not primary beam corrected,
but they were useful for measuring sizes of the compact components.}.
In the sub-mm we obtained 
JCMT SCUBA\footnote{http://www3.cadc-ccda.hia-iha.nrc-cnrc.gc.ca/jcmt/}
images at 450 and 850 $\mu$m
and recent CSO BOLOCAM\footnote{http://irsa.ipac.caltech.edu/data/BOLOCAM\_GPS/}
images at 1.1 mm \citep{ba+10}.
In the mid-infrared we obtained
MSX\footnote{http://irsa.ipac.caltech.edu/applications/MSX/MSX/}
images at 8, 12, 15 and 21 $\mu$m, Spitzer 
MIPS\footnote{http://irsa.ipac.caltech.edu/applications/Spitzer/Spitzer/}
images at 24 and 70 $\mu$m and IRAS reprocessed 
HIRES\footnote{http://irsa.ipac.caltech.edu/IRASdocs/hires\_over.html}
images at 60 and 100 $\mu$m. The recent paper \citep{ba+10} discusses the BOLOCAM   
1.1 mm data, and shows the Spitzer 24 $\mu$m as a greyscale in their Figure 18
with 1.1 mm contours. They also show 350 $\mu$m CSO SHARC-II data, but do not
quote fluxes (and we did not have access to the 350 $\mu$m from public 
archives).

The extra flux density data are summarised in Table \ref{tab:SED_extras}.
The limits for Sgr B2 (N) at 70, 24 and 21 $\mu$m are 3-sigma upper limits
(see later discussion).

\begin{table}
\caption{Fitted flux densities used for the Sgr B2 spectral energy 
distribution, in addition to values from the literature.}
\label{tab:SED_extras}
\begin{tabular}{cccccc}
\hline
Frequency   & Telescope  & \multicolumn{4}{c}{Flux density} \\
or          & Project or & Total & N      &  M     &  S   \\
Wavelength  & Instrument & Jy    &  Jy    &  Jy    &  Jy     \\
\hline
1.54 GHz    & ATCA     &        &  2.35  &  2.57 &  0.13 \\
9.1 GHz     & VLA      &        &  6.3   &  7.9  &  0.64 \\
36 GHz      & ATCA     &        &  6.1   &  8.0  &       \\
1.2 mm      & SIMBA    &  330   & 103    &  71   &  3.7  \\     
1.1 mm      & BOLOCAM  &  420   &  64    &  90   &  25   \\
0.85 mm     & SCUBA    &  2500  & 330    & 260   &  22   \\
0.45 mm     & SCUBA    & 68000  & 3200   & 3300  & 720   \\   
100 $\mu$m  & IRAS     & 65000  &        &       &       \\
 60 $\mu$m  & IRAS     & 14500  &        &       &       \\
 70 $\mu$m  & Spitzer  &        & $<$ 280  &       &       \\
 24 $\mu$m  & Spitzer  &        & $<$ 0.40 &       &       \\
 21 $\mu$m  & MSX      &        & $<$ 2.1  &  16.4 &  9.5  \\
 15 $\mu$m  & MSX      &        &        &  1.80 &  2.13 \\
 12 $\mu$m  & MSX      &        &        &  1.07 &  0.96 \\
  8 $\mu$m  & MSX      &        &        &  0.79 &  0.83 \\
\hline
\end{tabular}
\end{table}

The flux density data (Fig. \ref{fig:SEDs}) show two physical 
components, the dust peak around 1.5 THz (200 $\mu$m) and the 
optically thin free-free emission between 3 and 100 GHz.
There are, however, problems in using flux densities obtained from data
obtained with a range of spatial resolutions. For the fitting of the complex
structure with the components Sgr B2 (N), (M) and (N), and total flux,
different amounts of flux will be included depending on the size of the 
telescope beam (as the four components are a simplification). 
This leads to discrepancies in the flux densities,
even considering the quoted uncertainties. We have fitted models to the
SEDs, despite this, but note that the reduced $\chi^{2}$ of the fits are high.

For the free-free component, we have fitted power laws over the frequency
range up to around 100 GHz, excluding some points for Sgr B2 (N) and the total
which show some upturn due to the dust component. For the dust component we 
fitted the 3-parameter model discussed above. 
For Sgr B2 (N), as we discuss below, we also follow \cite{go+93} in including an 
extra absorption term for dust along the line of sight, making a 4-parameter fit
$ S_{\nu} = \Omega B_{\nu}(T) (1 - \exp(-\tau_{\nu,1})) \exp(-\tau_{\nu,2}) $.
The dust fit used the data around 100 GHz for Sgr B2 (N) and the total, after 
correction for the free-free component.
The fitting was done in the 
{\sc gnuplot} package, which uses the nonlinear least-squares (NLLS) 
Marquardt-Levenberg algorithm. 
For the total flux, we fitted the data after subtracting the models for
the compact components Sgr B2 (N), (M) and (S), so that we were actually fitting
the model to the extended envelope. The model fits for both the total flux
and the extended envelope are both plotted in Fig. \ref{fig:SEDs}, and we
note that for most of the frequency range, the extended component dominates
the total flux.

\begin{table*}
\caption{Parameters for the Sgr~B2 components fitted from the SEDs, and derived
from these fits. From the fit to the optically-thin free-free part of each 
spectrum, we obtain the spectral index $\alpha_{ff}$ and the flux density $S_{30}$ 
at reference frequency 30 GHz. We choose to fix the angular size of the dust
component to that measured from the sub-millimetre images, and derive the dust
temperature $T_{d}$ and turnover wavelength $\lambda_{\tau = 1}$ from the dust 
peak in the spectrum. The total hydrogen column density $N_{H}$, number density
$n_{H}$ and mass $M_{H}$ are derived from the dust spectrum fits, as described 
in the text.}
\label{tab:SED_fits}
\begin{tabular}{ccccccccccc}
\hline
Component & $\alpha_{ff}$ & $S_{30}$ & $\Omega$  & Diameter & $T_{d}$ & 
$\lambda_{\tau = 1}$ & $N_{H}$ & $n_{H}$ & $M_{H}$ \\
          &   &   &  fixed      &       &     &          & 
10$^{24}$ cm$^{-2}$ & 10$^{6}$ cm$^{-3}$ &  \\
          &   &  Jy  & arcsec$^2$ &  pc &  K   &  $\mu$m  & 
(10$^{28}$ m$^{-2}$)  & (10$^{12}$ m$^{-3}$) & M$_{\odot}$ \\
\hline
N        & -0.02 $\pm$ 0.03 &  7.1 & 117 & 0.40  & 65 $\pm$ 17 &  740 $\pm$ 150 &  
12   & 8.9   &   8600  \\
M        &  0.22 $\pm$ 0.02 &  9.9 & 138 & 0.43  & 50 $\pm$  1 &  680 $\pm$ 70 & 
9.6  & 6.8   &   8400  \\
S        &  0.04 $\pm$ 0.16 & 0.62 & 118 & 0.40  & 48 $\pm$  1 &  230 $\pm$ 30 & 
 1.1  & 0.87  &    850  \\
envelope & -0.32 $\pm$ 0.12 & 17.7 & 17400 & 4.5 & 24 $\pm$  2 &  280 $\pm$ 40 & 
 3.0  & 0.19  &  320000 \\
\hline
\end{tabular}
\end{table*}

For the free-free component, we find spectral indices of
$-0.02 \pm 0.03$, $0.22 \pm 0.02$, $0.04 \pm 0.16$ and $-0.32 \pm 0.12$
for the components Sgr~B2 (N), (M), (S) and the extended envelope respectively,
and flux densities at 30 GHz 7.1 Jy, 9.9 Jy, 0.62 Jy and 17.7 Jy respectively.
Note that this indicates that around half the flux is in the extended component
which may be missed by the interferometric observations (e.g. Fig 
\ref{fig:RRL_7mm_overlay}).
We do attribute the radio emission from 3 to 100 GHz in the compact components
to optically thin free-free emission, with basically a flat spectrum, although 
the rising spectrum of Sgr~B2 (M) may indicate some of the subcomponents
are ultracompact H II regions with the turnover from optically thick to 
optically thin above 10 GHz. The spectrum of the total Sgr~B2 flux is flat
(spectral index $-0.17 \pm 0.07$), but when we subtract the compact components,
notably  Sgr~B2 (M) with the rising spectrum, the residual extended component
has a somewhat steeper spectrum.

There has been some discussion of non-thermal radio emission from
Sgr~B2 by \citet{ho+07}, \citet{cr+07} and \citet{pr+08}.
We agree with \citet*{lapago08} and \citet{pr+08}
that due to the complex structure of Sgr~B2,
it is difficult to measure the radio spectral index accurately by combining
observations with different resolutions which may be including different
features. In particular, the fluxes measured with the GBT by \citet{ho+07}
for Sgr~B2 (N) will include flux from the extended component, to a greater
extent at the lower frequencies with the larger beam.

The fits to the dust components, for three free parameters $T$, $\Omega$ and 
$\lambda_{\tau = 1}$ gave similar results to that of \citet{go+93}, as expected
since the data used here is based on their compilation. However, there are
quite large uncertainties in derived $\Omega$ and $\lambda_{\tau = 1}$,
and the two parameters are highly correlated. We decided, therefore, to
constrain the fits by using a fixed derived value for the dust 
solid angle $\Omega$
obtained using the extra spatial information from the sub-mm images.
This fit to two free parameters ($\Omega$ and 
$\lambda_{\tau = 1}$) gives similar reduced $\chi^2$ to the three free 
parameter fits, and only changes the fitted temperature $T$ by a few K.

We used the geometric mean of the deconvolved sizes from the SCUBA 450- and 
850-$\mu$m images (resolution 7.5 and 14 arcsec), and the 1.3-mm IRAM 30-m
image (resolution 11 arcsec), giving gaussian half-widths 10.2, 11.2 and 10.2
arcsec for Sgr~B2 (N), (M) and (S). \footnote{We did not include deconvolved 
sizes from the BOLOCAM 1.1-mm or SIMBA 1.2-mm images, as the images were 
lower resolution (33 and
24 arcsec) and hence the deconvolved sizes were much less accurate.} 
We therefore used solid angles $\Omega = 1.133 \theta^{2}$ of 117, 138 and 
118~arcsec$^2$ for Sgr~B2 (N), (M) and (S). For the extended component, we used
the geometric mean of effective fitted solid angles from the SCUBA 450- and 
850-$\mu$m, BOLOCAM 1.1-mm and  1.3-mm IRAM-30-m images, which is
$\Omega = 17400$~arcsec$^2$. \footnote{The SIMBA 1.2-mm image may suffer from 
spatial filtering in the data reduction, affecting the extended structure.}

The results from the dust spectrum fits, with $\Omega$ fixed, and two free 
parameters $T$, and $\lambda_{\tau = 1}$ are given in Table \ref{tab:SED_fits}.

For Sgr~B2~(N) we find, as pointed out by \citet{go+93}, that the extra
line-of-sight absorption component to the model is needed, making in this 
case then a three parameter fit.  
If we fix the angular size $\Omega$ at that measured, then the
extra absorption component to the model is clearly required: the reduced 
$\chi^2$ of the fit without the absorption (7.7) is substantially greater than 
with the absorption included (4.4), and the fit is clearly poor for the sub-mm 
range.

The turnover
wavelength is fitted as $\lambda_{\tau = 1} = 100 \pm 17 \mu$m, in good agreement
with \citet{go+93}. This is physically interpreted as  Sgr~B2~(N) being partly
embedded in the envelope. The fit shown in Fig. \ref{fig:SEDs} does not use
the upper limits as constraints, but the non-detection of Sgr B2 (N) in the
Spitzer MIPS 24 micron data (at $<$ 0.40 Jy) is consistent with the fit, and is 
a strong constraint. The fit does rely strongly on the 100 and 50 micron points 
of \citet{go+92}, quoted in \citet{go+93} where Sgr B2 (N) was detected 
as a feature somewhat merged with the stronger peak of Sgr B2 (M) and the 
extended envelope. Our fitting of the recent Spitzer MIPS 70 micron image
shows Sgr B2 (M) and (S) saturated, so that these components and the
envelope cannot easily be fitted. The upper limit we quote for Sgr B2 (N)
at 70 micron is the flux at the known (from sub-mm) position, in
addition to some extended structure, which indicates that the 50 micron
flux from \citet{go+92} may be an over-estimate, including some flux from 
extended structure. This would indicate more absorption (and longer
$\lambda_{\tau = 1}$) may be appropriate, but better high resolution data
in the far-infrared range are required to confirm this. 

Due to the inter-relationship
of parameters $T$, $\Omega$ and $\lambda_{\tau = 1}$ it is possible to fit
the cutoff in the dust SED at high frequencies (infra-red, where Sgr~B2~(N) is 
not detected, see Fig. \ref{fig:radio_submm_ir}) with angular size $\Omega$ 
as a free parameter, which gives a low temperature 
($T$ around 30 K), but this requires a large angular size 
$\Omega \sim 500$~arcsec$^{2}$ which is in disagreement with the measured 
angular size. In addition, if Sgr~B2~(N) had this low temperature, similar
to that of the extended envelope, (and there was not the absorption), then the 
70 micron 
Spitzer MIPS image would show a clear peak at the Sgr B2 (N) position,
in addition to the extended structure. This is not the case, which is further
evidence against the low temperature interpretation of the Sgr B2 (N) SED. 

We also list in Table \ref{tab:SED_fits} the parameters derived from the dust
spectrum fits of total hydrogen column density $N_{H}$, number density
$n_{H}$ and mass $M_{H}$ for the four components Sgr~B2 (N), (M), (S)
and envelope. Note that $N_{H} = N(H\,I) + 2 N(H_{2})$ so this includes
both molecular and atomic hydrogen.
The turnover wavelength $\lambda_{\tau = 1}$ is
related to total hydrogen column density $N_{H}$ by
$ N_{H} =  1.4 \times 10^{20} \rm{cm}^{-2}
(\lambda_{\tau = 1}/\mu\rm{m})^{2} (Z_{\sun}/Z)/b $
where $Z/Z_{\sun}$ is the metallicity, assumed 2 for the Galactic Centre,
and $b$ is a dust parameter, assumed 3.4 for components Sgr~B2 (N), (M) and (S),
with $n_{H} \ge 10^{-6} \rm{cm}^{-6}$ and 1.9 for the envelope with
$n_{H} \le 10^{-6} \rm{cm}^{-6}$ \citep{go+93,mezywi90}. The number density
$n_{H}$ and mass $M_{H}$ are derived using the angular size $\Omega$ to estimate
the line-of-sight thickness, and hence volume of the components.
These are rough estimates, similar to those derived by \citet{go+93} from 
similar fits, but we assume Galactic Centre distance 8.0 kpc.

We note that the three hot cores Sgr~B2 (N), (M) and (S) have similar angular 
size and temperature, but that Sgr~B2 (S) has around an order of magnitude 
smaller flux density (both dust and free-free components), column density, 
number density and mass. The extended envelope is cooler but around an order of 
magnitude
larger in diameter than the cores, or two orders of magnitude in solid angle,
and dominates the flux density of both dust and free-free components.
The envelope is of much lower number density than the cores, but with 
much greater mass.

We use these parameters of the Sgr~B2 components in Section \ref{sec:discuss}
to interpret the molecular line observations of Section \ref{sec:results}.

\bsp

\label{lastpage}

\clearpage

\end{document}